\documentclass[acmlarge,screen]{acmart}

\usepackage{xcolor}
\usepackage{soul}
\usepackage{subcaption}
\usepackage{graphicx}
\usepackage{gensymb}
\usepackage{tabularx}

\AtBeginDocument{%
  \providecommand\BibTeX{{%
    \normalfont B\kern-0.5em{\scshape i\kern-0.25em b}\kern-0.8em\TeX}}}

\copyrightyear{2021}
\acmYear{2021}
\setcopyright{rightsretained}
\acmConference[MobileHCI '21]{Proceedings of the 23rd International Conference on Mobile Human-Computer Interaction}{September 27-October 1, 2021}{Toulouse \& Virtual, France}
\acmBooktitle{Proceedings of the 23rd International Conference on Mobile Human-Computer Interaction (MobileHCI '21), September 27-October 1, 2021, Toulouse \& Virtual, France}\acmDOI{10.1145/3447526.3472061}
\acmISBN{978-1-4503-8328-8/21/09}

\begin{document}

\title{Anticipatory Detection of Compulsive Body-focused
Repetitive Behaviors with Wearables}


\author{Benjamin Lucas Searle}
\affiliation{%
  \institution{University of Cambridge}
  \city{Cambridge}
  \country{UK}
}
\email{bls30@cam.ac.uk}

\author{Dimitris Spathis}
\affiliation{%
  \institution{University of Cambridge}
  \city{Cambridge}
  \country{UK}
}
\email{ds806@cam.ac.uk}

\author{Marios Constantinides}
\affiliation{%
  \institution{Nokia Bell Labs}
  \city{Cambridge}
  \country{UK}
}
\email{marios.constantinides@nokia-bell-labs.com}

\author{Daniele Quercia}
\affiliation{%
  \institution{Nokia Bell Labs}
  \city{Cambridge}
  \country{UK}
}
\email{daniele.quercia@nokia-bell-labs.com}

\author{Cecilia Mascolo}
\affiliation{%
  \institution{University of Cambridge}
  \city{Cambridge}
  \country{UK}
}
\email{cm542@cam.ac.uk}

\renewcommand{\shortauthors}{Searle et al.}
\renewcommand{\shorttitle}{Anticipatory Detection of Compulsive BFRBs with Wearables}

\begin{abstract}
Body-focused repetitive behaviors (BFRBs), like face-touching or skin-picking, are hand-driven behaviors which can damage one's appearance, if not identified early and treated. Technology for automatic detection is still under-explored, with few previous works being limited to wearables with single modalities (e.g., motion). Here, we propose a multi-sensory approach combining motion, orientation, and heart rate sensors to detect BFRBs. We conducted a feasibility study in which participants (N=10) were exposed to BFRBs-inducing tasks, and analyzed 380 mins of signals\footnote{Code and dataset are available: \url{https://github.com/Bhorda/BFRBAnticipationDataset}} under an extensive evaluation of sensing modalities, cross-validation methods, and observation windows. Our models achieved an AUC > 0.90 in distinguishing BFRBs, which were more evident in observation windows 5 mins prior to the behavior as opposed to 1-min ones. In a follow-up qualitative survey, we found that not only the timing of detection matters but also models need to be context-aware, when designing just-in-time interventions to prevent BFRBs.

\end{abstract}




\maketitle
\thispagestyle{empty}
\section{Introduction}
\label{sec:introduction}


The World Health Organization (WHO) defines mental health as our ability to function at a psychological level, characterized by autonomy, competence, independence, and actualization of emotional potential~\cite{who}. As WHO suggests, mental hygiene (the process of achieving mental health) could be compromised or put at risk, if certain symptoms or behaviors~\cite{molarius2009mental} are not diagnosed early and treated, both at an individual and collective level. For example, compulsive behaviors exhibit certain characteristics which may signal poor mental health. In literature, two terms that are interchangeably used to characterize these persistent behaviors are the Body-focused Compulsive Behaviors (BFCB) and the Body-focused Repetitive Behaviors (BFRB); for brevity, we use BFRB throughout the manuscript. These behaviors are repetitive in nature, primarily characterized by the use of hands, and exhibit distinctive behavioral signatures; face-touching, skin-picking, hair-pulling, to name a few, make up the BFRB list. These behaviors, if not identified early and corrected, may lead individuals to damage their physical appearance or, in extreme cases, to cause non-reversible physical damage to themselves~\cite{wilhelm1993nail,bohne2005pathologic}. But, more worryingly, they have been linked to the development of severe mental health problems~\cite{jmir1838}.

Behavioral studies leverage our digital traces~\cite{jiang2011target}, or our daily interactions with technology such as the use of smartphones or wearable devices~\cite{rachuri2010emotionsense, spathis19,tran2019modeling}. For example, Cherian et al.~\cite{cherian2017did} analyzed wrist-mounted accelerometer data to identify tooth brushing. In this work, by exploiting sensing capabilities of modern smartphones and wearables, we set out to understand behaviors related to our mental hygiene, not only a posteriori but also in anticipation\footnote{Throughout the paper, we use the terms ``prediction'' and ``anticipatory detection'' interchangeably.}; for example, in Competing Response Training\cite{competingresponse}, which is a prevalent treatment method, the prediction of these behaviors is crucial for a successful therapy outcome. By using Machine Learning (ML), we show how we can predict BFRBs early on from data preceding the behavior, so that successful interventions could be developed to prevent them. In so doing, we made three contributions:

\begin{itemize}
    \item We conducted, for the first time, a semi-controlled free-living experiment to study BFRBs, grounded on previous stress literature and taking into account the social context (\S\ref{sec:user_study}). Using consumer-grade wearables (i.e., Samsung Galaxy watches), we obtained 380 minutes of raw signals (i.e., accelerometer, gyroscope, and heart rate) from 10 users who underwent a series of BFRB-inducing tasks. We make our data set and pre-processing code publicly\footnote{\url{https://github.com/Bhorda/BFRBAnticipationDataset}} available to the community (\S\ref{sec:dataset}) in an anonymized manner.
    \item Using the collected data, we performed an in-depth analysis to investigate the feasibility of predicting compulsive behavior using a multisensory approach (\S\ref{sec:analysis}). Consistently across all models, we found that the combination of all sensor modalities yielded the best performance (\S\ref{sec:results}). In particular, we found that generic compulsive behavior (vs. normal) can be predicted with satisfactory discrimination of 0.89 AUC. In specific behaviors such as face touching and skin picking, our models achieved an AUC of 0.94. In addition, compulsive behaviors were more evident in an observation window of 5 minutes prior to the episode as opposed to 1-min window. The short observation windows allow for immediate interventions, while longer ones are more robust in terms of prediction accuracy due to leveraging the multi-sensory approach. Notably, the heart rate sensor was a stronger predictor in the longer window (taking into account that longer windows allow us to calculate richer heart rate variability metrics). On the other hand, the motion sensors dominated in 1-minute windows.
    \item In the light of these findings, we discuss the broader theoretical and practical implications of our work, given the recent re-emergence of face-touching as a public health risk for infectious diseases. Our system paves the way to just-in-time interventions for real-time compulsive behavior monitoring (\S\ref{sec:discussion}).
\end{itemize}

\section{Background and Related Work}
\label{sec:related_work}

\subsection{Body-focused repetitive behaviors}
\label{section:bfrb}
The term body-focused repetitive behavior (BFRBs), coined by Bohne~\cite{bohne2002}, refers to a set of impulsive behavioral disorders affecting a wide demographic. In this work, we focus on a subset of these behaviors, namely nail-biting, skin-picking, skin-biting, nail-picking, fidgeting, face-touching, and hair-pulling. This subset of behaviors is strongly correlated with clinically recognized body-dysmorphic conditions~\cite{keuthen2001skin}, and exhibit certain characteristics that can be uniquely identified from single source motion data.

Previous research suggests that the most common BFRBs are nail-biting with a prevalence of 34-64\%; skin-picking with a prevalence of 25\%; and hair pulling with a prevalence of 10.5\%~\cite{bohne2002}. This research was conducted on students from different cultures with comparable results on the prevalence of body dysmorphic disorder. These values suggest a sizeable demographic, further supporting the relevance of the current study. Understanding precise causes and triggers for these behaviors are essential for effective invocation in experiments. Relevant literature also suggests that real-time triggers are correlated to change in environment and, more importantly, stress~\cite{bohne2002}. Studies concluded, that the behaviors play an emotional regulation role, supported by findings showing these behaviors are triggered by and relieve impatience, boredom, and frustration~\cite{roberts2013emotion,roberts2015impact,williams2007function}. In an extended study, clinical approaches to handling emotion regulation are suggested as a method of treatment for BFRBs~\cite{roberts2016role}, with one study showing the pervasive treatment of depression to result in a decrease of body dysmorphic behavior~\cite{jmir1838}. This correlation between stress and BFRBs inspired the multisensory approach of our study: to combine stress monitoring through the proxy of heart-rate data, as well as motion data for a more accurate preemptive signal for BFRB occurrences.

Current literature suggests treatment of these conditions includes competing response training (CRT)~\cite{competingresponse}, habit-reversal, and cognitive therapy~\cite{wilhelm1993nail,lochner2017excoriation}. An example of CRT for nail biting or other habits involving the hands, pertains to holding the hands down at the side and making a fist or grasping objects \cite{MILTENBERGER2002911}. CRT consists of two sequential phases: the first concerns the identification of behavioral occurrences, while the second relies on countering the behavior with a competing response in which the behavior cannot be carried out. Our work helps in the first stage of CRT, by assisting the user in the recognition of when compulsive behaviors may occur, incorporating methods from the area of human activity recognition (HAR). 

\subsection{Human Activity Recognition for compulsive behaviors}
Most HAR approaches follow a common ML pipeline Lara~\cite{har}: data gathering, data processing, feature extraction, and model training. The most common data sources are motion data, environmental data, physiological data, and location data \cite{bulling2014tutorial, liu2009uwave}. We consider our task a specific case of gesture recognition \cite{caramiaux2014adaptive}.
Wrist-worn devices have been shown to help accurate inference of multiple conditions and states using ML methods~\cite{tison2018passive,rachuri2010emotionsense,ballinger2018deepheart, laput2019sensing, parate2014risq}. Previous attempts have been made to detect BFRBs with wearables, however, there is no existing research on the prediction of these behaviors. Azaria~\cite{azaria2016thumbs} developed Thumbs-Up, a wrist-worn device, which is used to infer hand-to-mouth behavior with a 92\% accuracy. Lu~\cite{lu2019detection} produced a device for habit detection using a wrist-worn device and machine learning algorithms. Several commercial wearables have also been devised to combat these conditions including the Tingle \cite{son2019thermal}, Keen\footnote{\url{https://habitaware.com/}}, and Pavlok\footnote{\url{https://pavlok.com}} devices. However, among this list, only the Tingle device has published peer-reviewed documentation of their results and proven applicability to BFRB monitoring. This device uses the same sensors (such as in our method) in addition to thermal sensors to detect proximity and position using a long short-term memory (LSTM) neural network. However, all the above devices provide in-time notifications to the wearer when a BFRB-esque motion is detected, in contrast to our approach focusing on the prediction.

\section{Research goals}
\label{sec:researchgoals}
As stated in the previous section, most prior work in BFRB monitoring relied on subjects seeking medical professional advice, as well as systematic surveys. Additionally, due to their repetitive nature, these behaviors may differ in free-living conditions as opposed to controlled laboratory settings. 
Therefore, we set out to understand \emph{whether these behaviours could be predicted using wearable-sensing data and, if so, how much in advance this could be achieved.} To address that, we subsequently formulated three research questions as follows:

\begin{enumerate}
    \item[\textbf{RQ\textsubscript{1}}:] Which features among gyroscope, accelerometer, and heart rate are good predictors of BFRBs?
    \item[\textbf{RQ\textsubscript{2}:}] What window sizes prior to BFRBs occurrences allow for reliable anticipation?
    \item[\textbf{RQ\textsubscript{3}:}] To what extent does a multisensory approach (motion and heart rate) improves BFRB prediction? 
\end{enumerate}

We study them in two phases. In phase I, we conducted a semi-controlled free-living experiment in which 10 subjects underwent a series of BFRB-inducing tasks, while we collected motion and heart rate data from a wearable device. In phase II, we extensively modeled the collected data using cross-validation methods, observation windows, and machine learning classifiers.

\section{User Study}
\label{sec:user_study}
In phase I, we conducted a semi-controlled free-living experiment to collect BFRB behavioral traces that we used in phase II to model these behaviors. Here, we describe the experimental procedure and, subsequently (\S\ref{sec:dataset}), present the collected data which we made publicly available in an anonymized manner\footnote{\url{https://github.com/Bhorda/BFRBAnticipationDataset}}.

\begin{figure*}
\centering

\begin{subfigure}[a]{0.95\textwidth}
   \includegraphics[width=1\linewidth]{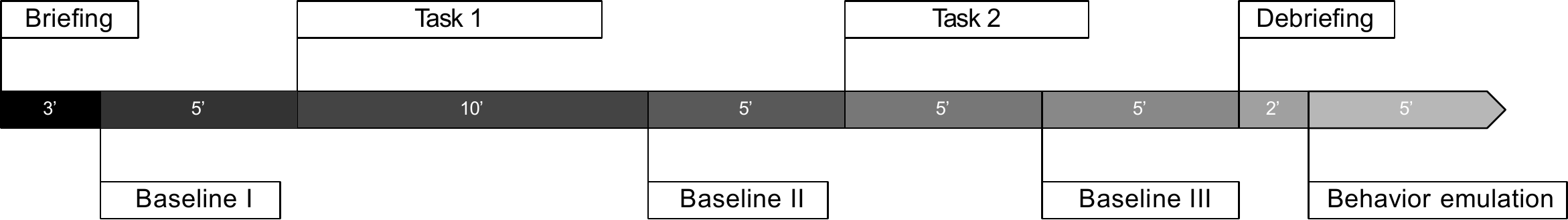}
   \caption{}
\end{subfigure}

\begin{subfigure}[b]{0.95\textwidth}
   \includegraphics[width=1\linewidth]{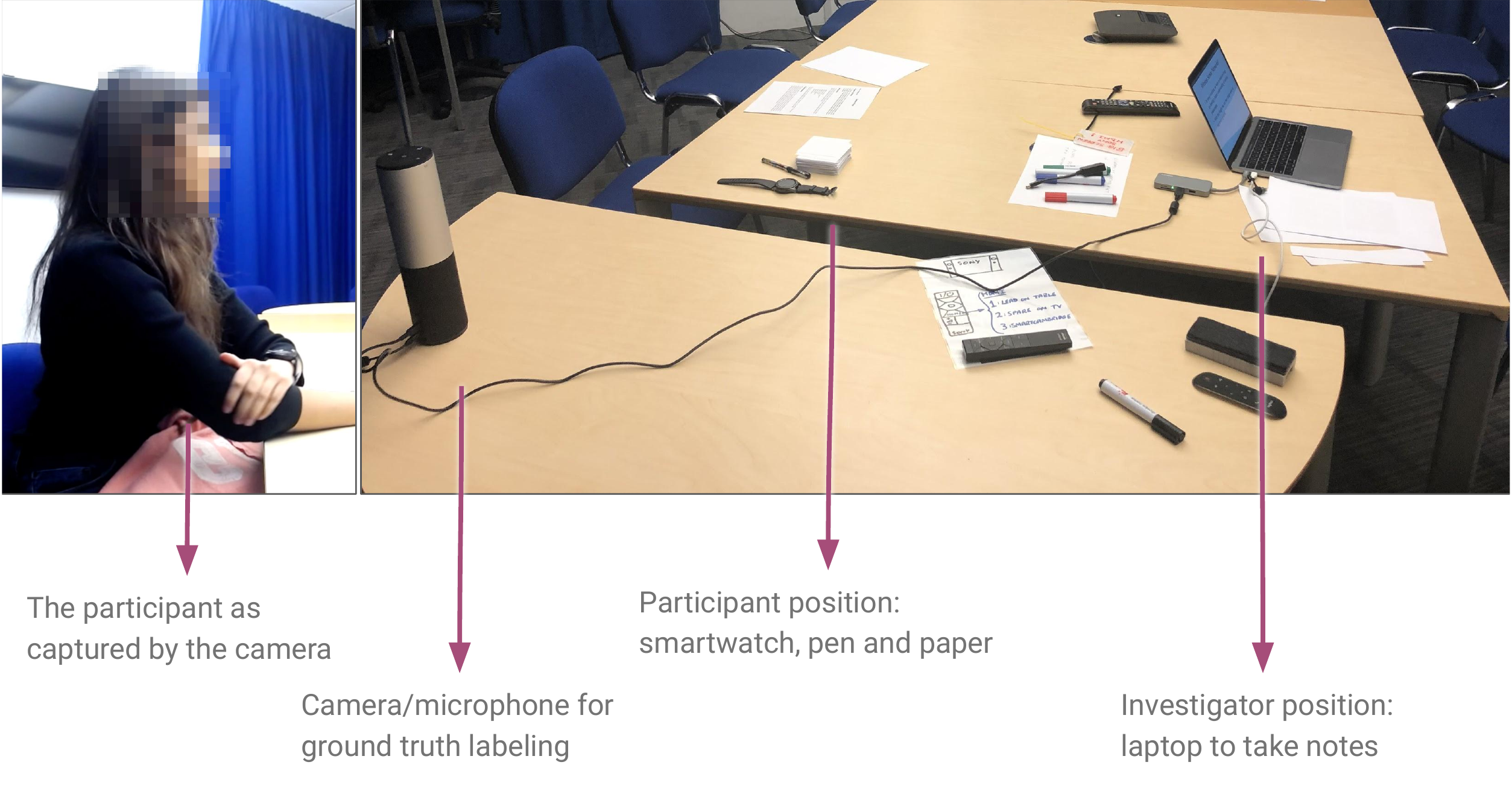}
   \caption{}
\end{subfigure}

\caption[]{\textbf{User study setup. } Experiment timeline \textbf{(a)}: Baseline resting periods (sitting idle) and tasks (\emph{Task 1:} presentation and \emph{Task 2:} arithmetic test) that induced stress and, eventually, triggered compulsive behaviors.  Experimental setup \textbf{(b)}: Subjects wore a smartwatch throughout the trial and completed stress-inducing tasks.}
\label{fig:experiment}
\end{figure*}

\begin{figure}
    \centering
    \includegraphics[width=0.9\textwidth]{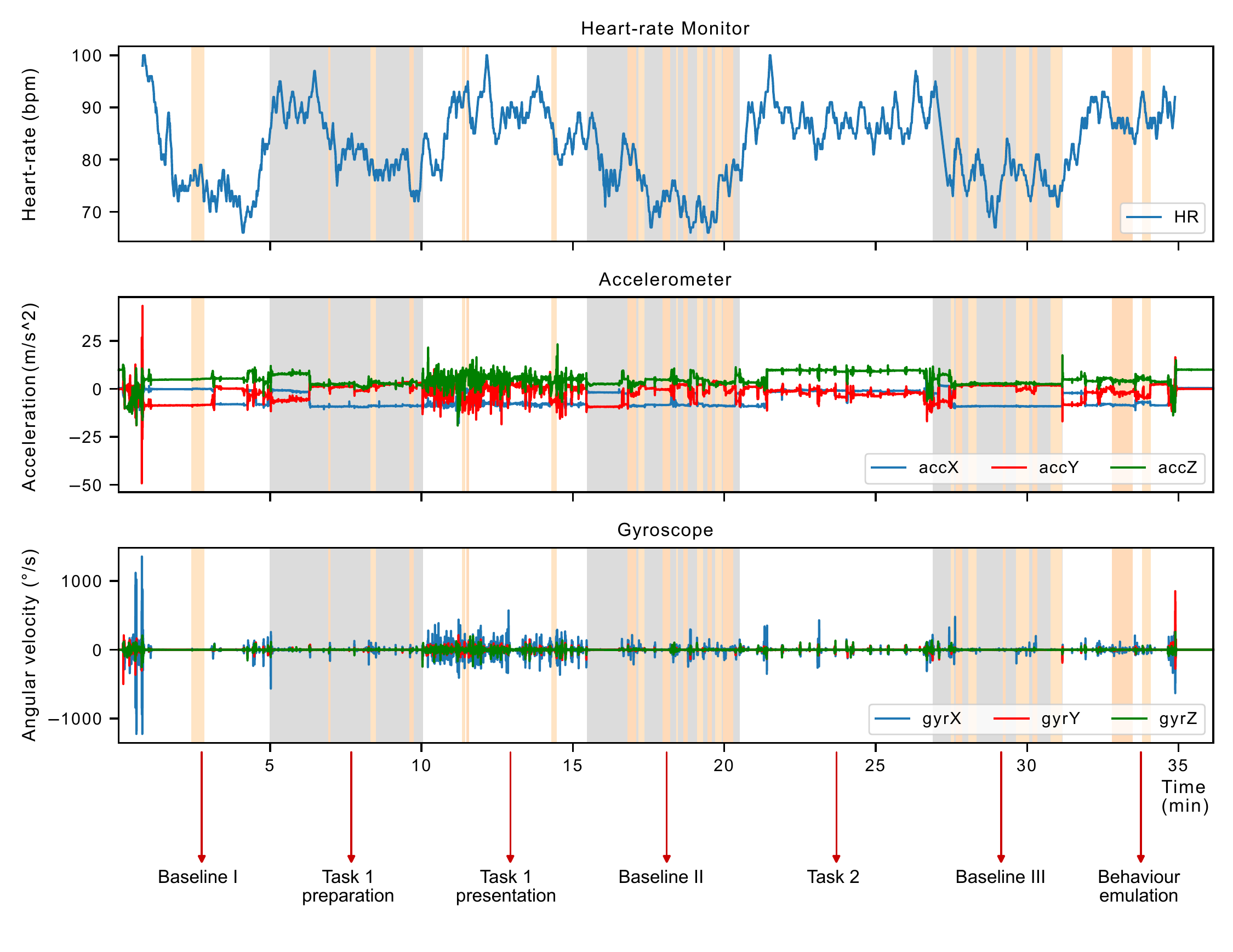}
    \caption{\textbf{Example of raw signals along with labels and experiment stages.} The orange-yellow highlights mark observed compulsive behaviors. The alternating white and light-gray segments mark the stages of the experiment. Data from Participant \#3.}
    \label{fig:exampledata}
  \end{figure}

 
\subsection{Participants}
We recruited 10 participants (4 female, 6 male), aged between 18 and 40 (M = 23.6, SD = 5.4), equally split between undergraduate students and early-career researchers, and with no prior history of any cardiovascular disease. Additionally, all participants were instructed not to consume coffee on the day of the experiment. The study was approved from the Ethics Committee of University of Cambridge. During the recruitment process, we informed them in writing that the experiment aimed at studying participants' behavior in a series of tasks. Note that, BFRBs were not mentioned to participants until after the experiment, during the last stage of debriefing and behavior emulation. This deliberate decision allowed us to eliminate any potential biases introduced before and during the experiment.

\subsection{Materials and Apparatus}
To carry out the experiment, we used the following apparatus and materials (Figure \ref{fig:experiment} a): sheet of paper for task 1, pen, \textasciitilde 9 sheets of paper for task 2, a room with few distractions (ideally large empty space), a device for recording audio and video, and a Samsung Galaxy smartwatch application that we developed for collecting accelerometer, gyroscope, and heart-rate data. We acknowledge that other sensors that measure different body characteristics such as galvanic skin response--electrodermal activity (GSR/EDA), skin temperature, or breathing rate could complement our study \cite{taylor2017personalized}. However, we opted for HR/HRV and motion for two reasons: (i) skin sensors are linked to physiological arousal contrary to HRV which is deemed a better proxy to the autonomic nervous system (and in turn to stress) \cite{schmidt2019wearable}, and, (ii) commercial consumer-grade wearables offer greater support for HR/HRV over GSR/EDA (e.g., most smartwatches have a PPG sensor compared to the expensive Empatica E4 for GSR).


To induce BFRBs, we used the Trier Social Stress test~\cite{kirschbaum1993trier} for invoking moderate social stress, founded on the correlations between BFRBs and stress~\cite{bohne2002}. 
While there are numerous protocols aiming at eliciting stress in studies involving human subjects \cite{schmidt2019wearable}, we opted for a well-studied and frequently employed stress elicitation protocol, that is, the Trier Social Stress test. The protocol suggests a 10-minute anticipation period, followed by a 10-minute presentation period and, finally, by a 5-minute arithmetic task. To obtain participants' behavior traces, we used an application for Samsung watches~\cite{wellbeat}. The application continuously records motion and physiological readings through the device's sensors at 10Hz sampling rate.


\subsection{Study Protocol and Procedure}



Modifications were made to the original Trier's test protocol to accommodate the opportunity to perform compulsive behaviors in the form of additional baseline periods of inactivity before and after the tasks, where participants were asked to remain silent and otherwise inactive. Consequently, one of these periods separated the two tasks instead of having a single test interval. The arithmetic task (i.e., Task 2) was also modified to avoid speaking, to combat possible confounding with heart rate increase due to, for example, vocal tension. We posit that real-world stress induced by the presentation and arithmetic tasks could generalize to other social situations and we consider them as a suitable set up for out study.   \par

\begin{table}[]
\centering
\small
\begin{tabular}{llll}
\hline
\multicolumn{2}{l}{\textbf{Modalities}}      & \textbf{Definition}                                                                                &  \\ \hline
\multicolumn{2}{l}{\textbf{Motion}} &                                                                                                    &  \\
               & accX               & The X-axis of the accelerometer data (horizontal movement)                                         &  \\
               & accY               & The Y-axis of the accelerometer data (vertical movement)                                           &  \\
               & accZ               & The Z-axis of the accelerometer data (depth movement)                                              &  \\
               & gyrX               & The X-axis of the gyroscope data (roll angle)                                                      &  \\
               & gyrY               & The Y-axis of the gyroscope data (yaw angle)                                                       &  \\
               & gyrZ               & The Z-axis of the gyroscope data (pitch angle)                                                     &  \\ \hline
\multicolumn{2}{l}{\textbf{Heart}}  &                                                                                                    &  \\
               & HR                 & The instantaneous heart rate in BPM                                                                &  \\
               & RMSSD              & Proxy for heart rate variability defined as the root mean square of successive differences of interbeat intervals
 &  \\ \hline
               &                    &                                            & 
\end{tabular}
\caption{\textbf{Feature categories included in the data set.} Each category consists of 4 features, namely mean (mean), standard deviation (std), minimum (min), and maximum (max). In later figures, features are denoted as `sensor + feature name' such as `accXstd'-referring to the standard deviation of the accelerometer data in the X axis.}
    \label{table:features}
\end{table}

The experiment included two tasks which induced BFRBs and three resting periods in between. These tasks were selected due to their potency to induce social stress and, thus BFRBs. Other tasks with similar properties could necessarily be used in future work. There was a briefing at the beginning of the experiment, and a debriefing followed by behavior emulation at the end of the experiment (Figure ~\ref{fig:experiment}b). The experiment lasted around 35 minutes and audiovisual recording was used throughout the experiment. This allowed us to obtain ground truth data for modeling purposes (\S\ref{sec:analysis}). \par

At the beginning of the study, participants were instructed to wear the watch on their non-dominant hand, as it is the most common position and would not introduce motion artefacts during the writing task. There was no further instruction in what to do with their non-dominant hand. After equipping the apparatus, the experiment started with a 5-minute baseline period, followed by the first task. The first task was to prepare and present a 5-minute presentation of the participant's curriculum vitae (resume) with pen and paper provided for notes. After the preparation period (stage 2. in Figure~\ref{fig:experiment}b), the notes and pen were confiscated and the participant was asked to give the presentation (stage 3. in Figure~\ref{fig:experiment}b). If the participant had run out of material they were asked to continue talking until the time limit was up. The task aimed to invoke \emph{anticipatory stress} in the first phase and \emph{performance stress} in the second phase. This was followed by another baseline period of 5-minutes, after which the second task was concluded. The second task involved an arithmetic task and required participants to write down as many powers of 3 as possible starting from $3^0$ up to $3^{15}$. Every 30 seconds, the participant was asked to discard their work and restart on a new piece of paper. This task attempted to invoke \emph{frustrative stress}. There was a final baseline period following this, after which the participant was debriefed and asked to emulate any compulsive behavior they were aware of exhibiting in their day-to-day life. This simulated data was not utilized later in the study and only served as initial exploration of the data. Timestamps describing the experiment structure were recorded by the investigator throughout the study, and used as a reference during the data preparation phase (\S\ref{sec:dataset}). An example timeline of the data collected can be seen in Figure \ref{fig:exampledata}.

\section{Data set}
\label{sec:dataset}

\subsection{Data processing pipeline}



\begin{figure}
    \centering
    \includegraphics[width=0.9\textwidth]{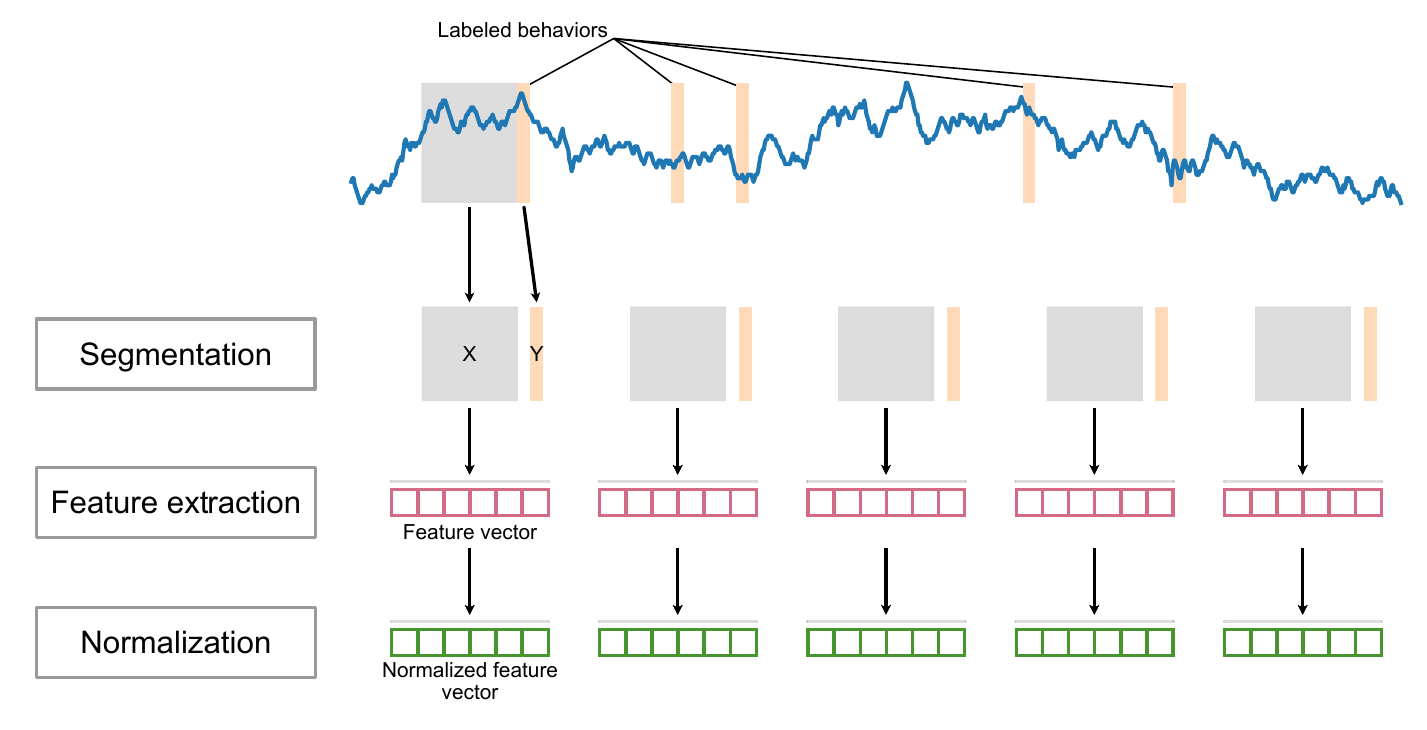}
    \caption{\textbf{Data Preparation Pipeline.} Stages of data formatting to prepare the data sets for model training. Data was labeled according to the recorded footage of the experiments. Data was normalized to the first Baseline period as described in \S \ref{sec:user_study}. In the segmentation phase, the y windows mark the labeled behaviors, before which the x windows are generated depending on the window size. For the feature extraction, only the x window is taken into account.}
    \label{fig:pipeline}
  \end{figure}

Having obtained a data set of a total of 380 minutes of raw motion and heart rate signals, we developed a 5-stage pipeline to process it. The data processing pipeline is comprised of: data labeling, data segmentation, feature extraction, and data normalization (Figure \ref{fig:pipeline}), which we describe next. The entire pipeline is modular with flags for normalization technique, unique or aggregate analysis of participants, specific or aggregate analysis of different compulsive behaviors, and custom segment sizes. All data manipulation was performed using the Pandas and NumPy libraries for Python.


\vspace{6pt}\noindent\textbf{Labeling.} 
To derive the ground truth labels, we made use of the audiovisual recordings. We labeled the time delta from the start of the recording when a compulsive behavior was observed. Additionally, we marked which hand was involved in the compulsive behaviors to allow for different combinations for analysis of motion data.


\vspace{6pt}\noindent\textbf{Segmentation.}
This set of procedures ensured that the output is in the correct format for machine learning. The first segments of the data are cut into fixed-length chunks of time series from each sensor (referred to as windows). Each window has two components, namely `x', the series of data points before the start of the observed compulsive behavior (referred to as x-windows), and `y', the series of data points starting from the time of behavior observations (referred to as y-windows). The input for the module is the desired length of the x-window and the y-window (Figure~\ref{fig:pipeline}). We note that this process differs from a canonical rolling window approach because our label of interest is not distributed uniformly in the data. Instead, we first isolate the compulsive behavior `y', and then use a window just before it occurred `x'. In inference (prediction) time, the models would just require the input window `x', which is a process that can be executed continuously on-device. We further differentiated between two types of windows:
\begin{itemize}
    \item \textbf{Positive windows:} These windows include observed compulsive behaviors in the y-window. Positive windows were tagged with the type of behavior they mark, and whether they are `\textbf{clean}' or `\textbf{dirty}'. Clean positive windows refer to those with no recorded compulsive data in the x-window, whereas the `dirty' descriptor refers to windows with compulsive behavior observed in the x-window.
    \item \textbf{Negative windows:} These windows were selected randomly from the data set where no recorded compulsive behaviors were observed in the y-window. The number of negative windows generated was equal to the number of positive windows per data set; this ensures the feature set is balanced.
\end{itemize}
It is worth noting that the overlap between windows is essentially guaranteed, as visible in Figure \ref{fig:exampledata}, due to the limitations of the duration of the study and the low average distance between observed behaviors. We will be using the notation of $Ax/By$ for the discussion of different size segments with `A' denoting the size of the x-window in seconds, and `B' denoting the size of the y-windows in seconds (e.g., $300x/1y$ would correspond to segment sizes of 300 seconds prior and 1 seconds post the start of the behavior).\par
In our study, two different segment sizes were examined, namely $60x/1y$, and $300x/1y$.  Prevalence of intra-participant correlations were significant, suggesting personalized training may yield better results. Segments with 3 second y-windows were also examined to exclude 1 and 2 second long behaviors in the data set where the behavior may not be prominent enough, however, this exposed no difference to the 1-second y-windows. Various different lengths of x-windows were also tested (2 to 4 min), which as we shall see in (\S\ref{sec:results}), results in significantly lower performance compared to 1- or 5-min windows. As stated in Laborde's work \cite{laborde2017heart}, the gold standard in conducting HRV analysis is a 5-min window (we did not include HRV in windows < 5 min). As these windows (< 5 min) do not include HRV, we expect them to rely more on motion than heart data, as with the 1-min windows. However, the motion data was noisier in these windows resulting in lower performance, thus confirming our initial hypothesis.\par

\vspace{6pt}\noindent\textbf{Feature extraction.}
Basic descriptive features were extracted from each x-window using the Pandas library. The descriptive data recorded includes the means, standard deviations, minimums, and maximums of each dimension of each sensor, namely the accelerometer, gyroscope, and heart-rate sensor. Additionally, using WellBeat's data processing pipeline~\cite{wellbeat}, we extracted Heart Rate Variability (HRV) parameters in 5-minute window data sets. In particular, we made use of a widely used HRV parameter, the RMSSD (i.e., the root mean square differences of successive RR intervals). Among the long list of HRV parameters~\cite{laborde2017heart}, we chose RMSSD as it has been found to be a good proxy for detecting stress. It is defined as the root mean square of the successive differences of R-R intervals, and it is computed as $ \sqrt{\frac{1}{n-1} \sum_{i=1}^{n-1}(RR_{i}-RR_{i+1})^2} $
, where n is the total number of RR intervals. We refer the reader to~\cite{wellbeat} for additional reading in conducting HRV analysis. 

\vspace{6pt}\noindent\textbf{Normalization.}We normalized all features using a z-score transformation as ($\frac{x - \mu_{b1}}{\sigma_{b1}}$), where $x$ is the value of each data point (Table~\ref{table:features}). This technique benefited from handling outliers better than min-max normalization, however, the output scale does not accurately represent the original data set. In our approach, the $\mu_{b1}$ and $\sigma_{b1}$ values were extracted from the first baseline period of each experiment (Figure~\ref{fig:experiment}b) as the physical and cognitive intensity of the study is not uniform. By using the overall scores, the output would be biased towards increases in each sensor. 

\subsection{Descriptive statistics}
\begin{figure}
    \centering
    \begin{subfigure}{.5\textwidth}
    \centering
    \includegraphics[width=1\textwidth]{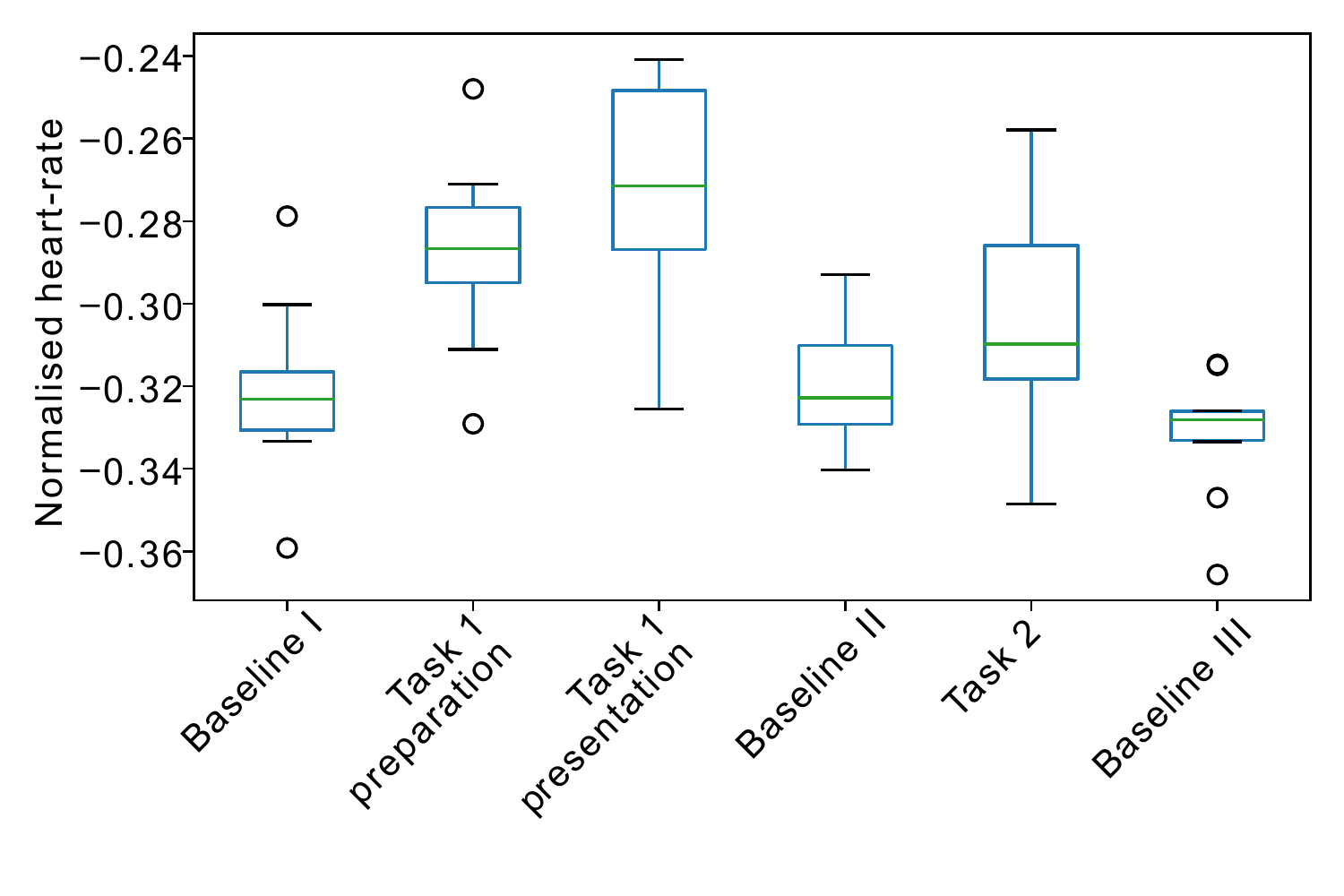}
    \caption{}
    \label{fig:hrmperstage}
    \end{subfigure}%
    \begin{subfigure}{.5\textwidth}
    \centering
     \includegraphics[width=1\textwidth]{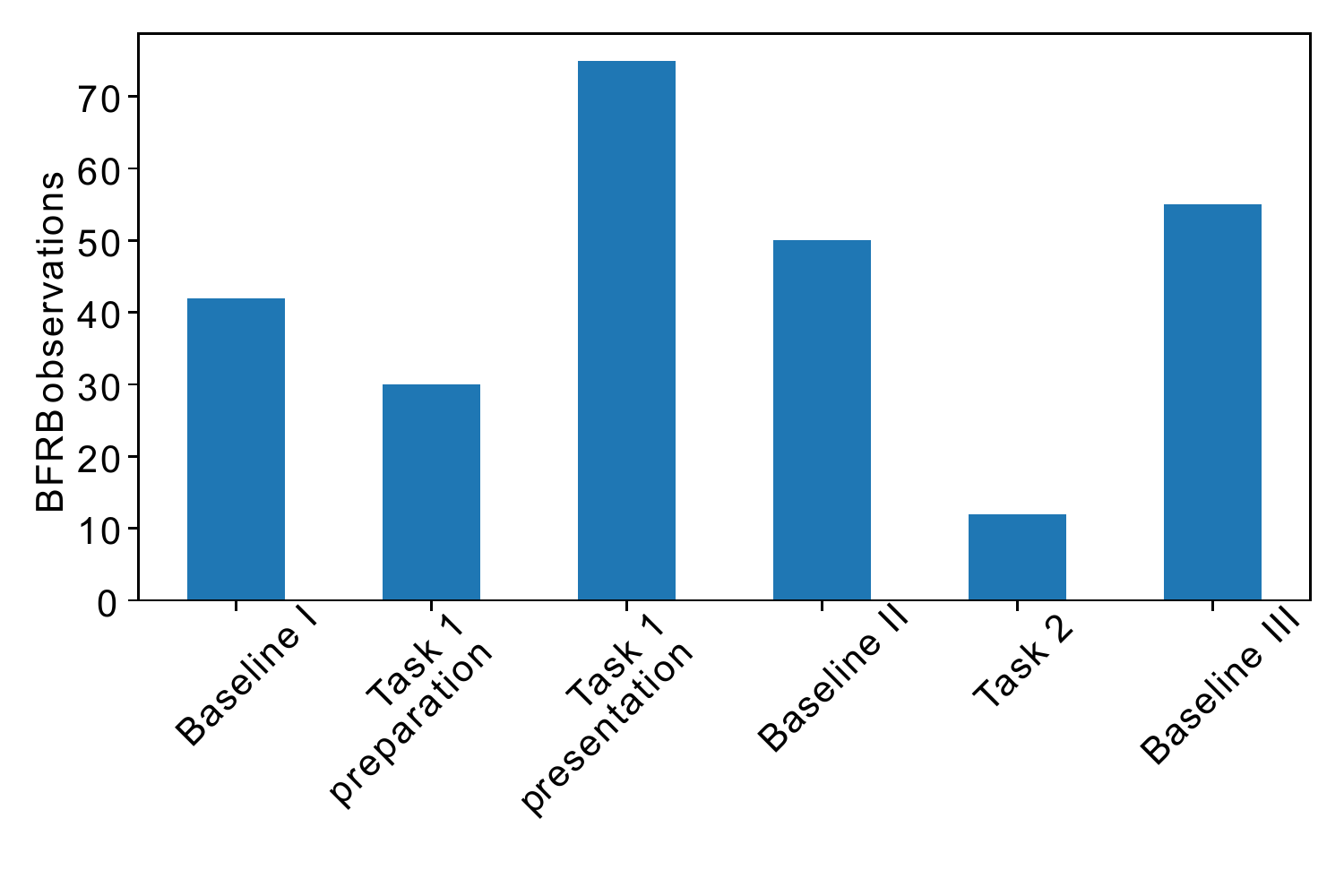}
    \caption{}
    \label{fig:cbperstage}
    \end{subfigure}
    \caption{\textbf{Descriptive statistics per stage.} Physiological and behavioral changes in the data per experiment stage. Normalized heart rates \textbf{(a)}: Normalized using z-score with $\mu$ and $\sigma$ from Baseline I. BFRB observation count \textbf{(b)}: Occurrences were reported with respect to the audiovisual recordings of experiments. }
\end{figure}

\begin{figure}
    \centering
    \begin{subfigure}{.5\textwidth}
    \centering
    \includegraphics[width=1\textwidth]{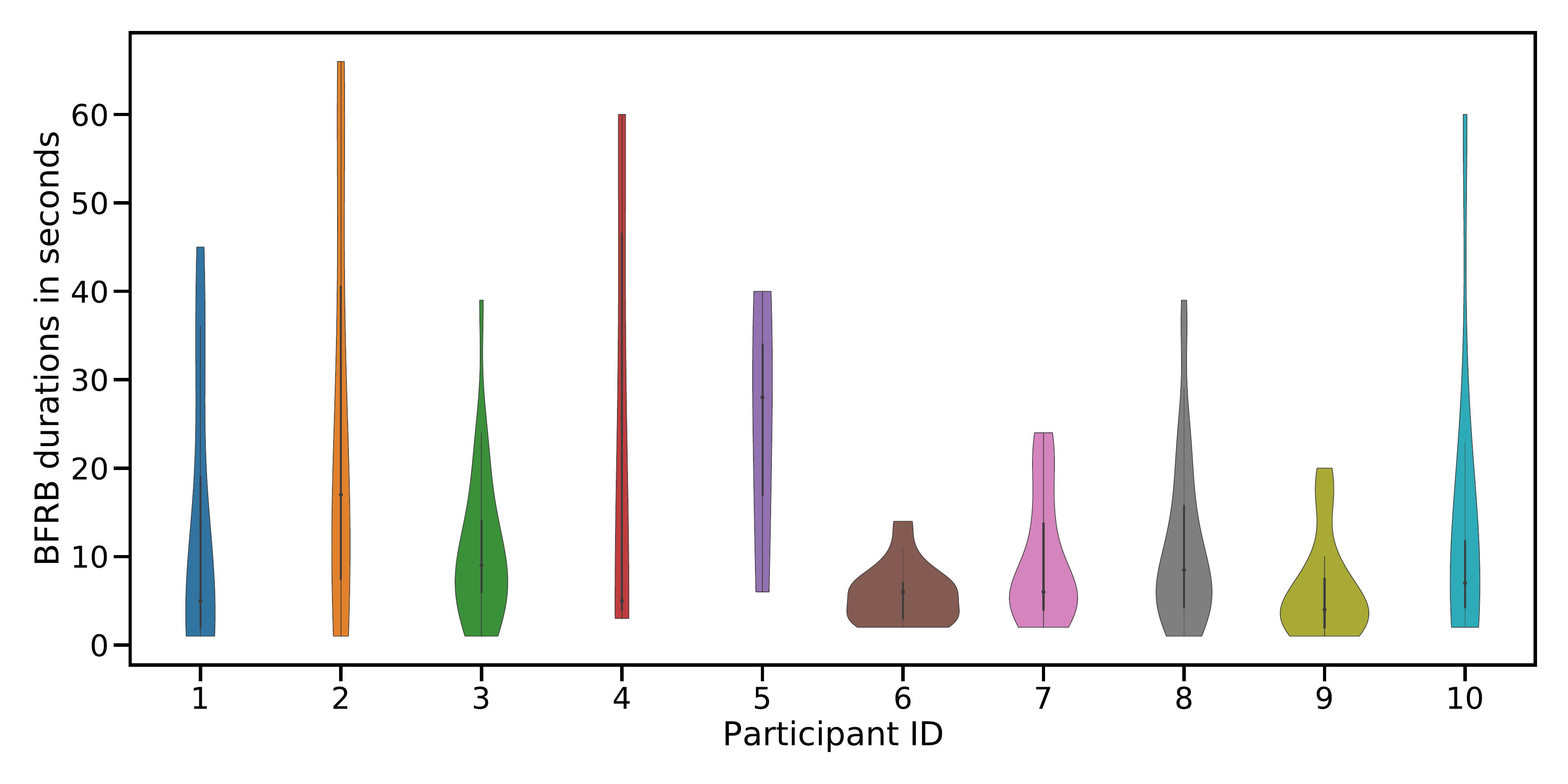}
    \caption{}
    \label{fig:durperparticipant}
    \end{subfigure}%
    \begin{subfigure}{.5\textwidth}
    \centering
     \includegraphics[width=1\textwidth]{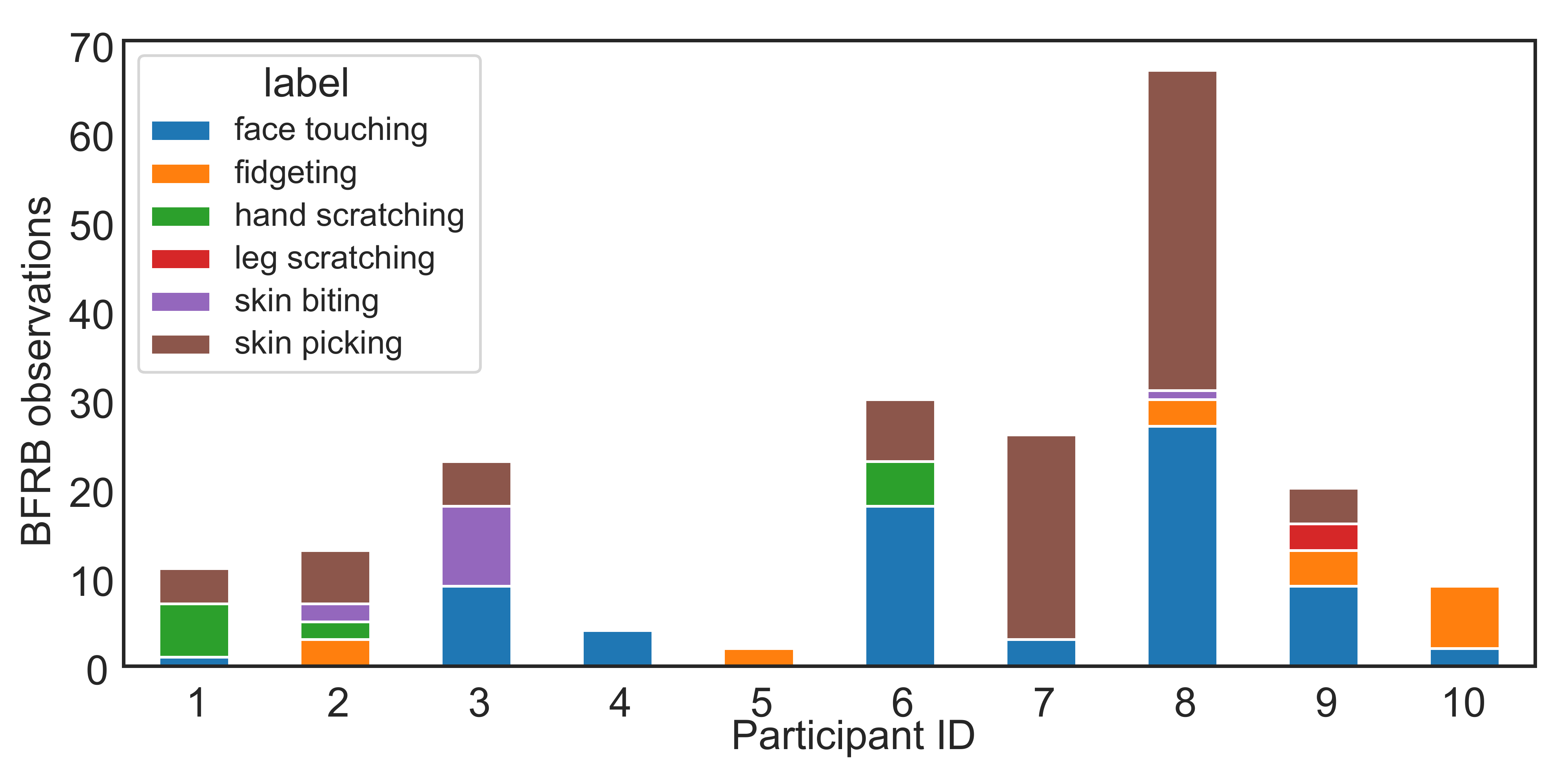}
    \caption{}
    \label{fig:cbperparticipant}
    \end{subfigure}
    \caption{\textbf{Descriptive statistics per participant.} Observed behavior durations and count per participant. Distribution of BFRB durations \textbf{(a)}: Start and end timestamps were reported from the audiovisual recordings of the experiments. BFRB observation count \textbf{(b)}: Behaviors were labeled by an investigator based on the recordings. }
\end{figure}


Whilst our findings in the prevalence of different behaviors do not align perfectly with previous research on the most common BFRBs (suggested to be nail-biting with a prevalence of 34-64\%; skin-picking with a prevalence of 25\%; and hair pulling with a prevalence of 10.5\%~\cite{bohne2002}), we saw the most prevalent behaviors to be skin-picking and face-touching. As we shall see in (\S\ref{subsec:survey}), 60\% of our participants admit to nail-biting in a follow-up survey; this corroborates the representativeness of our sample size. \par

Before moving into phase 2 of our study, we examined the hypothesis of whether stress would invoke compulsive behaviors. To do so, we compared the average heart rates during Task 1 and Task 2 to the baseline stages~\ref{fig:hrmperstage}. We then gathered the number of observed behaviors in each stage (Figure~\ref{fig:cbperstage}), and found that the average heart rate and the number of behaviors are indeed higher during Task 1 (presentation stage), concluding that there is a correlation between BFRBs and stress-as expected.



 
Figure~\ref{fig:cbperparticipant} and Figure~\ref{fig:durperparticipant} depict the distributions of behavior duration and the number of times compulsive behavior was observed, per participant. Note that only the 2 participants with the fewest BFRB observations (P4, P5) were observed displaying only one type of behavior. In total, skin-picking was the most prevalent behavior (41.2\%), followed by face touching (34.3\%), fidgeting (8.8\%), skin biting (6.5\%), hand scratching (6\%), nail-biting (1.4\%), leg scratching (1.4\%), and hair-pulling (0.4\%).





\section{Methodology}
\label{sec:analysis}

Using the collected data (\S\ref{sec:dataset}), in phase II, we set out to investigate whether these behaviors could be predicted and, the extent to which, they could be anticipated in advance. To address that, we conducted a series of experiments with various cross-validation methods, observation windows, and machine learning classifiers, which we describe next.

\subsection{Experiments}
We explored six experimental combinations along two dimensions: 
\begin{itemize}
    \item \textbf{Anticipatory window size}: Duration of activity included before observed behaviors. Analyzed window sizes are: \emph{(a)} 1-minute and \emph{(b)} 5-minute.
    \item \textbf{Label sets}: The type(s) of compulsive behavior included in the combination. Analyzed label sets are: \emph{(a)} all-compulsive behaviors, \emph{(b)} face touching, and \emph{(c)} skin picking. The latter two were selected due to their prevalence in the collected data (\S\ref{sec:dataset}).
\end{itemize}
Each combination relies on binary classification, with positive labels corresponding to data points describing the included labels of compulsive behaviors, and negative labels corresponding to periods of normal behavior.

\subsection{Classification and cross-validation methods}
We implemented and tested three classifiers: logistic regression (LR), random forest (RF), and gradient boosting trees (GBT). We included LR classifier as its low relative complexity makes it viable for mobile and offline training in addition to the interpretability of results. RF has been widely used in HAR tasks, while GBT is another state-of-the-art shallow learning algorithm, very well optimized with lower complexity than RF. 

To assess the performance of our classifier, we used standard classification performance metrics. These include the recall, the receiver operating characteristic (ROC), and the F1 score. We favored recall over precision due to the nature of our task as it deemed appropriate to predict the potential of compulsive behavior than to miss one. Put it differently, false positives cost less than false negatives in terms of diagnosis and prevention of these compulsive behaviors. The ROC plot shows the true positive rate against the false-positive one, and it is used to analyze and distinguish optimal models to sub-optimal ones. In our context, this allows gaining insights into which participants can be considered outliers in the data set, and to examine the contributions of different sensor modalities to the prediction. Finally, the F1 score shows a weighted average of precision and recall, and allows us to distinguish between the cost of false positives to false negatives.

In our experiments, we adopted two cross-validation strategies to model BFRBs similar to~\cite{constantinides2018personalized}. First, we explored a \emph{Leave-one-user-out} (generic) in which we excluded a participant from the test set and trained on the remaining users. Second, we explored a \emph{Participant-stratified} (personalized) in which we sampled 20\% from each participant's data set for the test set, and trained on the remaining data points. The personalized results should suggest the extent to which the models can be trained for individual users, whilst the generic method offers insight on the scalability to a greater demographic based on the smaller data set. 

\par
\section{Results}
\label{sec:results}

Tables~\ref{table:allgeneric} and~\ref{table:allpersonalized} report the results of the generic cross-validation and the personalized one, respectively. The results reflect high variance across cross-validation methods, label set, and anticipatory window size, with a significant increase in overall results using personalized cross-validation. All-compulsive results reflect the most consistent results across window sizes.


\begin{table}[]
    \resizebox{\textwidth}{!}{
        \begin{tabular}{rcccccc}
            \hline
            \multicolumn{1}{l}{}                        & \multicolumn{3}{c}{1-minute}                                       & \multicolumn{3}{c}{5-minute}                  \\
    \multicolumn{1}{l}{}                        & Recall        & AUC           & F1                                 & Recall        & AUC           & F1            \\ \hline
    \multicolumn{1}{l|}{\textbf{All-compulsive}}  &               &               & \multicolumn{1}{c|}{}              &               &               &               \\
    \multicolumn{1}{r|}{Logit Regression}       & 0.594 (0.326) & 0.725 (0.148) & \multicolumn{1}{c|}{0.581 (0.279)} & 0.699 (0.347) & \textbf{0.813} (0.238) & 0.698 (0.324) \\
    \multicolumn{1}{r|}{Random Forest}          & 0.515 (0.283) & 0.765 (0.100) & \multicolumn{1}{c|}{0.549 (0.257)} & 0.698 (0.309) & 0.663 (0.171) & 0.670 (0.219) \\
    \multicolumn{1}{r|}{Gradient Boost}         & 0.551 (0.297) & 0.740 (0.129) & \multicolumn{1}{c|}{0.555 (0.257)} & 0.851 (0.232) & 0.673 (0.226) & 0.749 (0.204) \\ \hline
    \multicolumn{1}{l|}{\textbf{Face touching}} &               &               & \multicolumn{1}{c|}{}              &               &               &               \\
    \multicolumn{1}{r|}{Logit Regression}       & 0.446 (0.377) & 0.413 (0.147) & \multicolumn{1}{c|}{0.363 (0.270)} & 1.000* (0.000) & \textbf{0.855} (0.205) & 0.845 (0.219) \\
    \multicolumn{1}{r|}{Random Forest}          & 0.203 (0.299) & 0.436 (0.160) & \multicolumn{1}{c|}{0.206 (0.256)} & 1.000* (0.000) & 0.540 (0.297) & 0.790 (0.141) \\
    \multicolumn{1}{r|}{Gradient Boost}         & 0.271 (0.221) & 0.356 (0.203) & \multicolumn{1}{c|}{0.293 (0.225)} & 0.940* (0.085) & 0.720 (0.311) & 0.815 (0.177) \\ \hline
    \multicolumn{1}{l|}{\textbf{Skin Picking}}  &               &               & \multicolumn{1}{c|}{}              &               &               &               \\
    \multicolumn{1}{r|}{Logit Regression}       & 0.266 (0.194) & 0.384 (0.357) & \multicolumn{1}{c|}{0.258 (0.181)} & 0.710 (0.418) & \textbf{0.852} (0.232) & 0.653 (0.350) \\
    \multicolumn{1}{r|}{Random Forest}          & 0.112 (0.148) & 0.220 (0.153) & \multicolumn{1}{c|}{0.134 (0.164)} & 0.745 (0.400) & 0.698 (0.320) & 0.620 (0.232) \\
    \multicolumn{1}{r|}{Gradient Boost}         & 0.378 (0.350) & 0.280 (0.163) & \multicolumn{1}{c|}{0.352 (0.180)} & 0.825 (0.271) & 0.622 (0.292) & 0.720 (0.103) \\ \hline
    \end{tabular}}
    \caption{\textbf{Means of generic cross-validation results.} Standard deviations are calculated over participant scores. \\
    *Around 60\% of participants' face touching data points were discriminated and excluded based on the missingness score of their HRV values. As stated in \cite{wellbeat}, to ensure reliable HRV estimates we filtered out scores <0.5, which led to a dropout of 11 labeled BFRB events belonging to a single participant. 
}
    \label{table:allgeneric}
    \vspace{2em}
    \resizebox{\textwidth}{!}{
    \begin{tabular}{rcccccc}
    \hline
    \multicolumn{1}{l}{}                        & \multicolumn{3}{c}{1-minute}                                       & \multicolumn{3}{c}{5-minute}                  \\
    \multicolumn{1}{l}{}                        & Recall        & AUC           & F1                                 & Recall        & AUC           & F1            \\ \hline
    \multicolumn{1}{l|}{\textbf{All-compulsive}}  &               &               & \multicolumn{1}{c|}{}              &               &               &               \\
    \multicolumn{1}{r|}{Logit Regression}       & 0.646 (0.124) & 0.720 (0.064) & \multicolumn{1}{c|}{0.658 (0.082)} & 0.762 (0.033) & 0.810 (0.059) & 0.802 (0.030) \\
    \multicolumn{1}{r|}{Random Forest}          & 0.828 (0.042) & \textbf{0.892} (0.011) & \multicolumn{1}{c|}{0.798 (0.019)} & 0.816 (0.056) & 0.808 (0.044) & 0.812 (0.045) \\
    \multicolumn{1}{r|}{Gradient Boost}         & 0.794 (0.122) & 0.862 (0.057) & \multicolumn{1}{c|}{0.784 (0.073)} & 0.862 (0.062) & 0.778 (0.032) & 0.836 (0.039) \\ \hline
    \multicolumn{1}{l|}{\textbf{Face touching}} &               &               & \multicolumn{1}{c|}{}              &               &               &               \\
    \multicolumn{1}{r|}{Logit Regression}       & 0.706 (0.115) & 0.634 (0.135) & \multicolumn{1}{c|}{0.614 (0.077)} & 0.914 (0.129) &\textbf{ 0.944} (0.077) & 0.914 (0.048) \\
    \multicolumn{1}{r|}{Random Forest}          & 0.704 (0.142) & 0.708 (0.090) & \multicolumn{1}{c|}{0.678 (0.101)} & 0.884 (0.159) & 0.884 (0.159) & 0.908 (0.075) \\
    \multicolumn{1}{r|}{Gradient Boost}         & 0.566 (0.068) & 0.612 (0.047) & \multicolumn{1}{c|}{0.576 (0.036)} & 0.916 (0.077) & 0.686 (0.343) & 0.904 (0.041) \\ \hline
    \multicolumn{1}{l|}{\textbf{Skin Picking}}  &               &               & \multicolumn{1}{c|}{}              &               &               &               \\
    \multicolumn{1}{r|}{Logit Regression}       & 0.514 (0.188) & 0.402 (0.091) & \multicolumn{1}{c|}{0.490 (0.113)} & 0.754 (0.112) & 0.804 (0.103) & 0.806 (0.071) \\
    \multicolumn{1}{r|}{Random Forest}          & 0.600 (0.273) & 0.614 (0.255) & \multicolumn{1}{c|}{0.554 (0.210)} & 0.968 (0.044) & \textbf{0.940} (0.040) & 0.940 (0.048) \\
    \multicolumn{1}{r|}{Gradient Boost}         & 0.684 (0.121) & 0.554 (0.099) & \multicolumn{1}{c|}{0.608 (0.092)} & 0.984 (0.036) & 0.912 (0.079) & 0.922 (0.053) \\ \hline
    \end{tabular}}
    \caption{\textbf{Means of personalized cross-validation results.} Standard deviations are calculated over 10 iterations with random seeds.}
    \label{table:allpersonalized}
\end{table}

\subsection{Best performing models}
\begin{figure}
    \centering
    \begin{subfigure}{.55\textwidth}
      \centering
    \includegraphics[width=\linewidth]{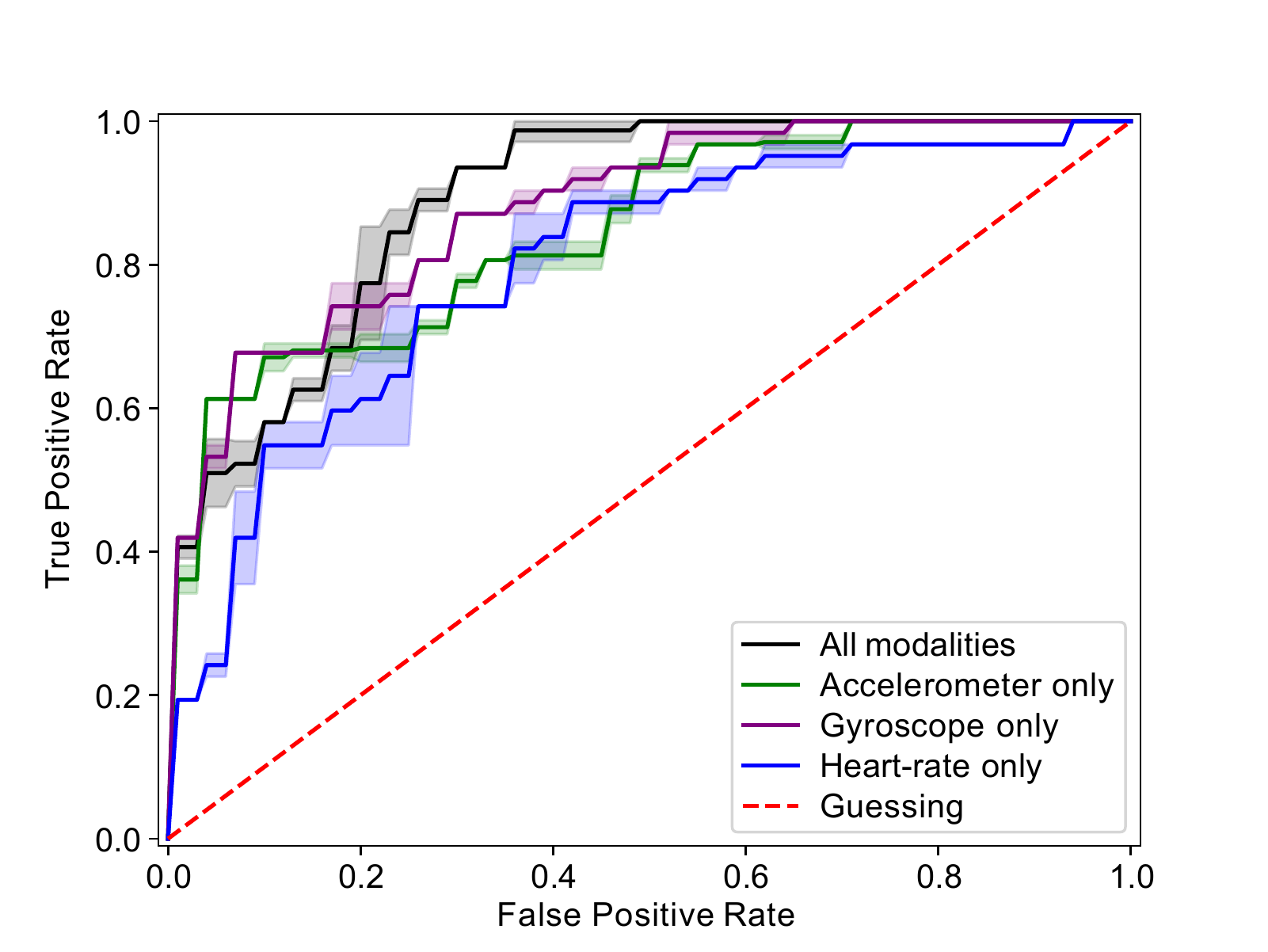}
      \caption{}
      \label{fig:bm1}
    \end{subfigure}%
    \begin{subfigure}{.5\textwidth}
      \centering
    \includegraphics[width=\linewidth]{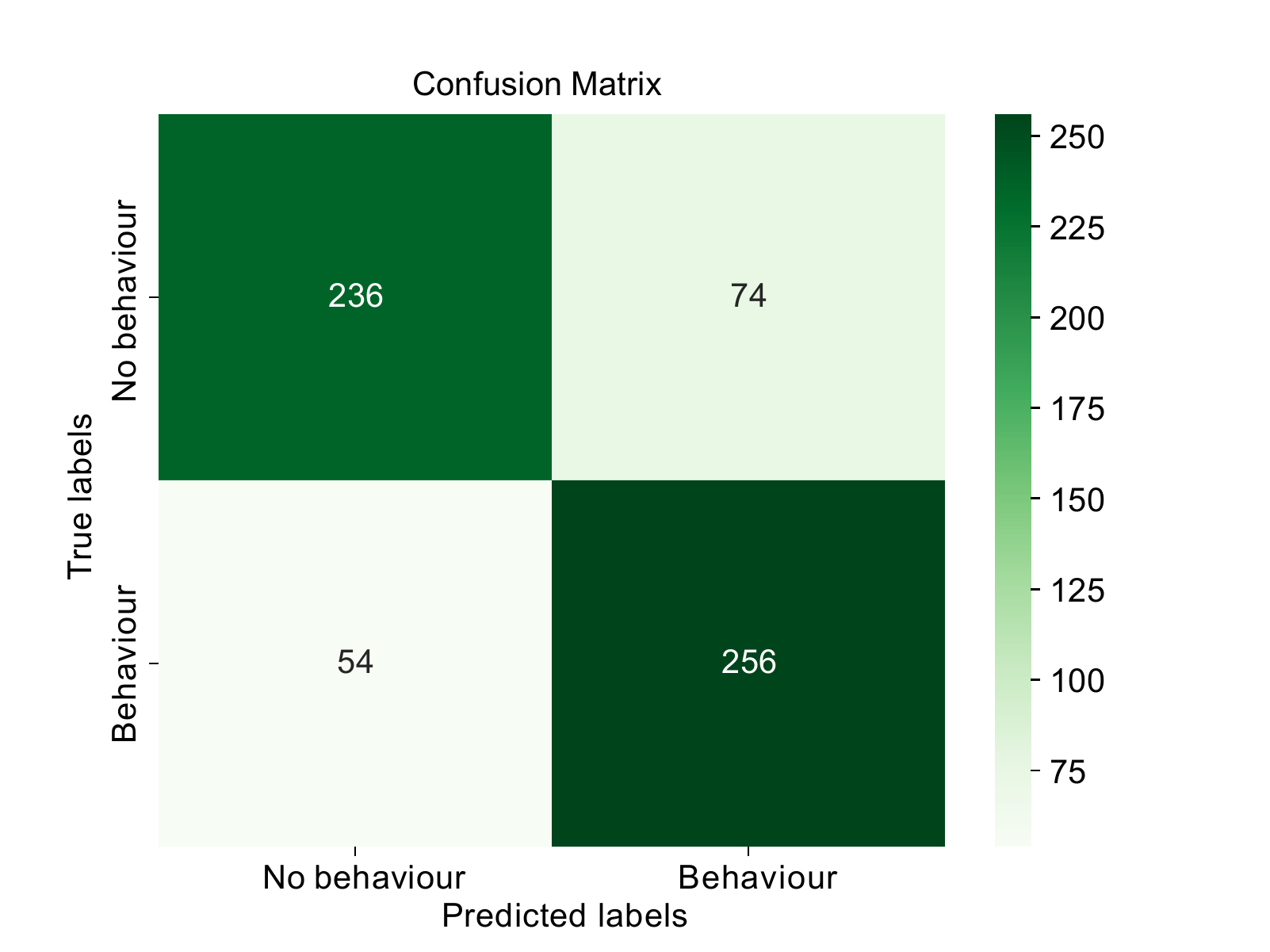}
      \caption{}
      \label{fig:bm2}
    \end{subfigure}
    \begin{subfigure}{.5\textwidth}
        \centering
        \includegraphics[width=\linewidth]{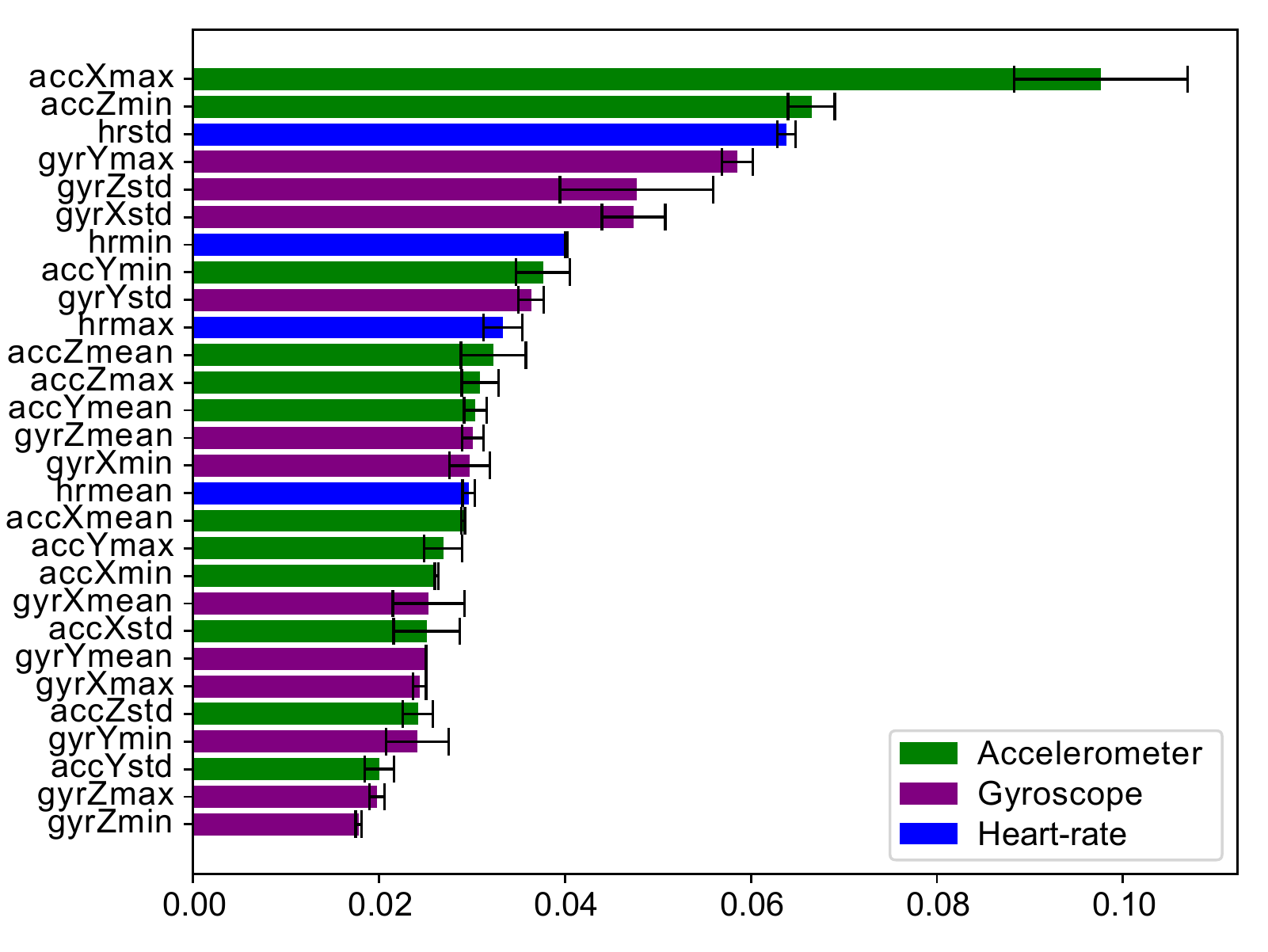}
        \caption{}
        \label{fig:bm3}
      \end{subfigure}%
      \begin{subfigure}{.5\textwidth}
        \centering
        \includegraphics[width=\linewidth]{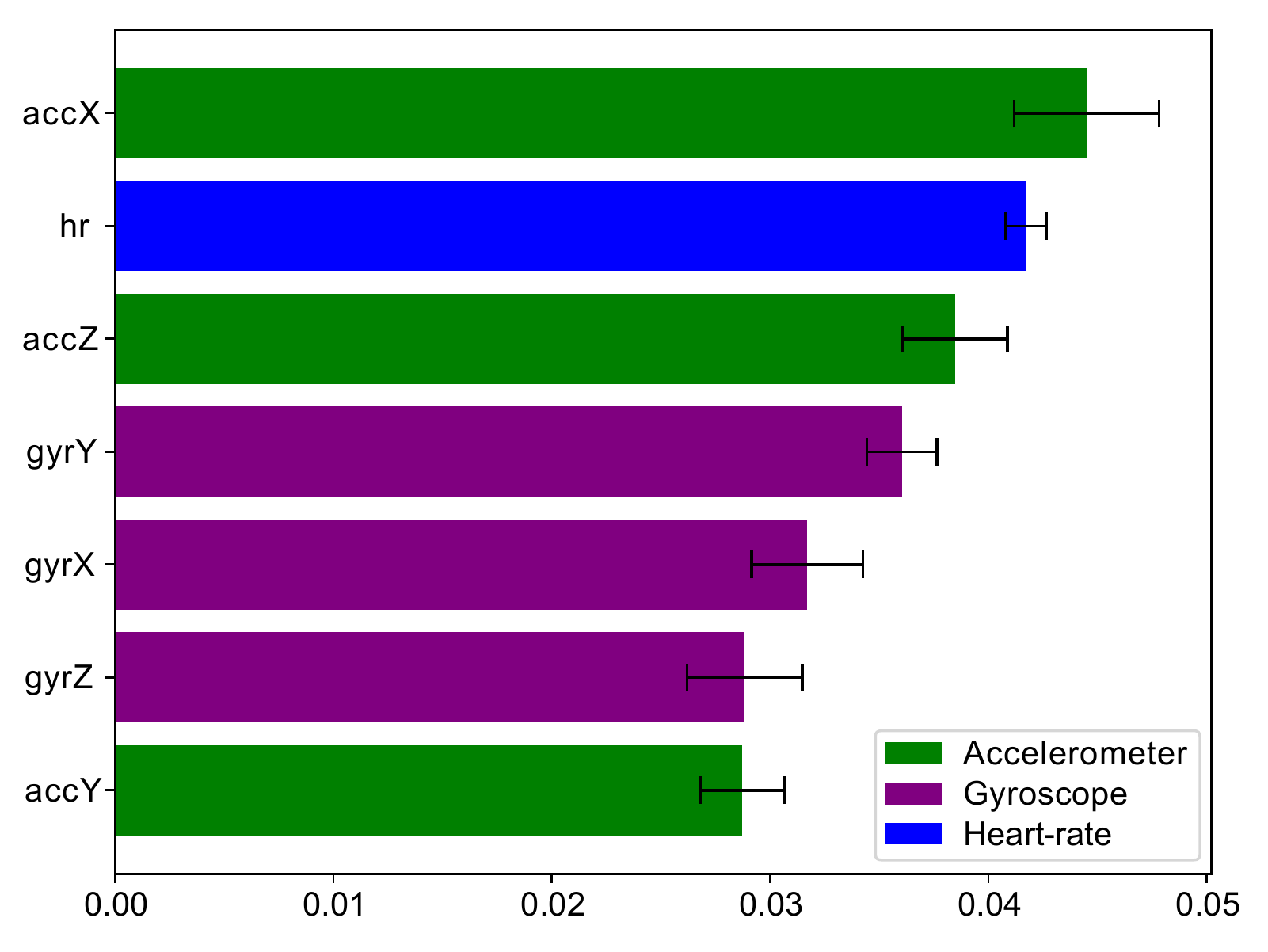}
        \caption{}
        \label{fig:bm4}
      \end{subfigure}
    \caption{\textbf{Results of participant-stratified 1-minute all-compulsive data using Random Forest.} \textbf{(a)} ROC-per-modality. \textbf{(b)} Confusion matrix. \textbf{(c)} All feature importances. \textbf{(d)} Features are grouped per modality. The notations refer to Figure \ref{table:features}.}
    \label{fig:bm}
\end{figure}

For brevity, we discuss our best performing model RF using the personalized cross-validation method. To identify the performance gain in the different modalities and, in turn, answer our \textbf{RQ\textsubscript{1}}, we analyzed the results and performed ablation tests (i.e., testing each modality in isolation). Figure~\ref{fig:bm} represents the entire analysis of the model. Figure~\ref{fig:bm1} shows that including all modalities increases overall reliability with a steady curve. The gyroscope data alone produces the best individual results, but suffers from having only a few strong predictors (Figure~\ref{fig:bm3}).

 

Heart rate ranks high in feature importance with standard deviation being the strongest performer from its features (Table \ref{table:features}). The model produced using gradient boosting performed best after this in the same task. Whilst the results are somewhat lower, the feature importance is largely the same, with the top 3 categories being accelerometer Z, accelerometer X, and heart-rate (Table \ref{table:features}). The disruption from the X-axis of the accelerometer can be noticed, however, the individual feature importances have the same strongest top 2, namely the accXmax and accZmin. The logistic regression model does not fare well in this comparison with the highest impact coefficients being those of the gyroscope, closely followed by heart-rate. The reliance of this model on linear interactions may be the cause of the lower performance.


\begin{figure}[!ht]
    \centering
    \begin{subfigure}[t]{.47\textwidth}
    \centering
\includegraphics[width=1\linewidth]{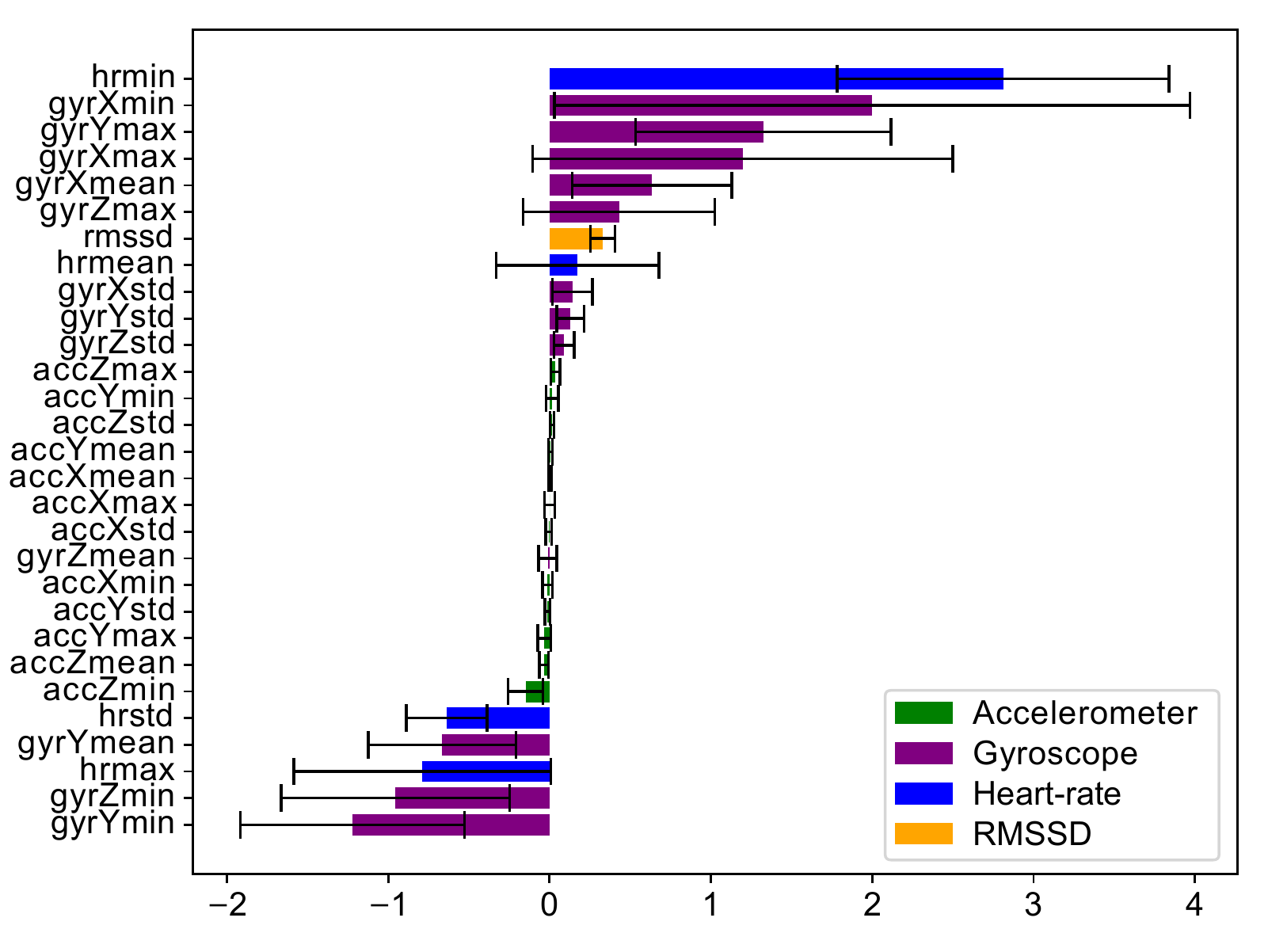}
\caption{}
\label{fig:fi}
\end{subfigure}%
\begin{subfigure}[t]{.53\textwidth}
    \centering
    \includegraphics[width=1\linewidth]{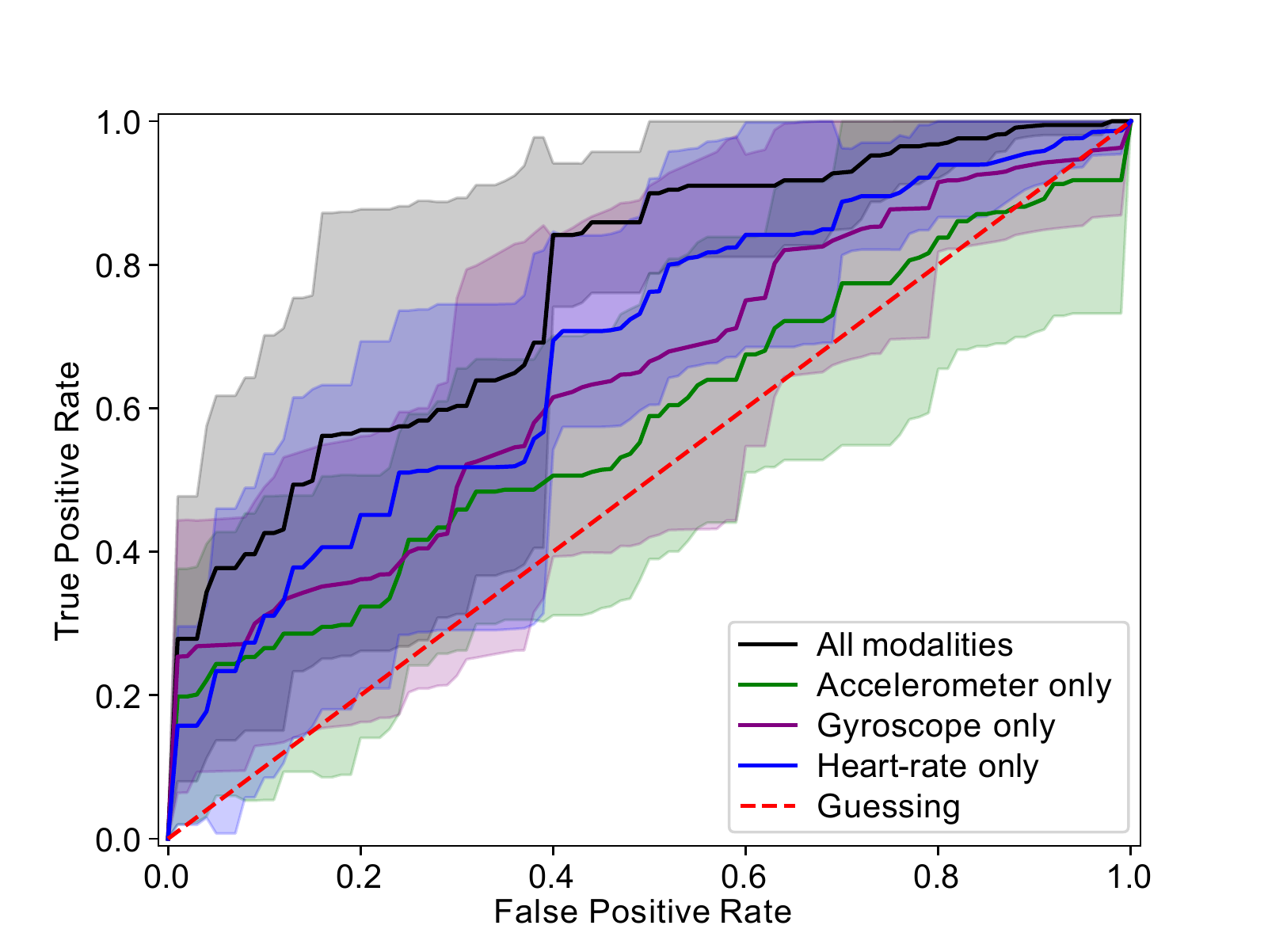}
    \caption{}
    \label{fig:rfpar}
    \end{subfigure}
    \caption{\textbf{Important predictors for generic cross-validation} \textbf{(a)} Coefficients per modality and axis in logistic regression model on the generic all-compulsive 5-minute window. \textbf{(b)} ROC curves of each modality in random forest model on generic all-compulsive 1-minute windows.}
\end{figure}

\subsection{Impact of cross-validation methods}
The results of the personalized cross-validation consistently outperform the generic one. We also observed lower standard deviations in the personalized results, with generic cross-validation resulting in significant imbalances between participants. Generic models relied heavily on heart-rate and HRV, whilst the personalized ones reflect the performance gain from resolving varying motion signatures between participants.  Overall, the findings from evaluating each cross-validation do not suggest that accurate prediction can be generalized but, by training a classifier to uniquely identify a single user's signature, the accuracy of prediction is increased significantly to the point of relevance.



\vspace{6pt}\noindent\textbf{Impact of anticipatory window sizes.}
We observed an overall performance increase in the 5-minute anticipatory window versus the 1-minute window in the isolated behaviors. In the gradient boosting and random forest models, this can be attributed to the additional RMSSD feature, providing better representation of heart-rate data when compared to instantaneous heart-rate. However, the best generic model is consistently logistic regression. Here, the model relies heavily on heart-rate and the gyroscope (Figure~\ref{fig:fi}). 

The personalized models tend to rely more heavily on motion data in the 1-minute windows, with heart-rate  ranking in the middle consistently in terms of feature importance; this is expected as the motion signature of these behaviors should be relatively similar. Generic classification shows a different activity, with 1-minute windows performing best on heart-rate alone when comparing ROC curves between the individual modalities seen in Figure \ref{fig:rfpar}, with 5-minute windows following a similar pattern with heart-rate being replaced with RMSSD. These results shed light on our \textbf{RQ\textsubscript{3}}, allowing us to conclude that the signature of heart activity is indeed indicative of compulsive behaviors, with richer features allowing for an increase in accuracy (Figure~\ref{fig:bm1}), and corroborating our multisensory approach. Furthermore, better prediction of a smaller subset of prevalent compulsive behaviors can be achieved using cheaper sensors, such as the accelerometer and gyroscope. With respect to \textbf{RQ\textsubscript{2}}, we conclude that both 1-minute and 5-minute anticipatory window sizes are sufficiently accurate for BFRBs prediction. However, these behaviors were more predictable consistently across most experiments when using 5-minute windows.

\vspace{6pt}\noindent\textbf{Effect of the granularity of behavior types.}
By analyzing behaviors separately, we did not find any significant increase in the performance compared to all-compulsive predictions (Tables \ref{table:allgeneric} and \ref{table:allpersonalized}). Whilst the performance of 5-minute windows equipped with HRV were sufficient, the motion data signature of behaviors could not be generalized over the population. Personalized validation did show an increase in all configurations of face touching and skin picking results over their generic counterparts, supporting the benefits of personalized models for BFRBs inference using motion data.


\subsection{Follow-up survey}
\label{subsec:survey}

\begin{figure}
    \centering
    \includegraphics[width=1.04\textwidth]{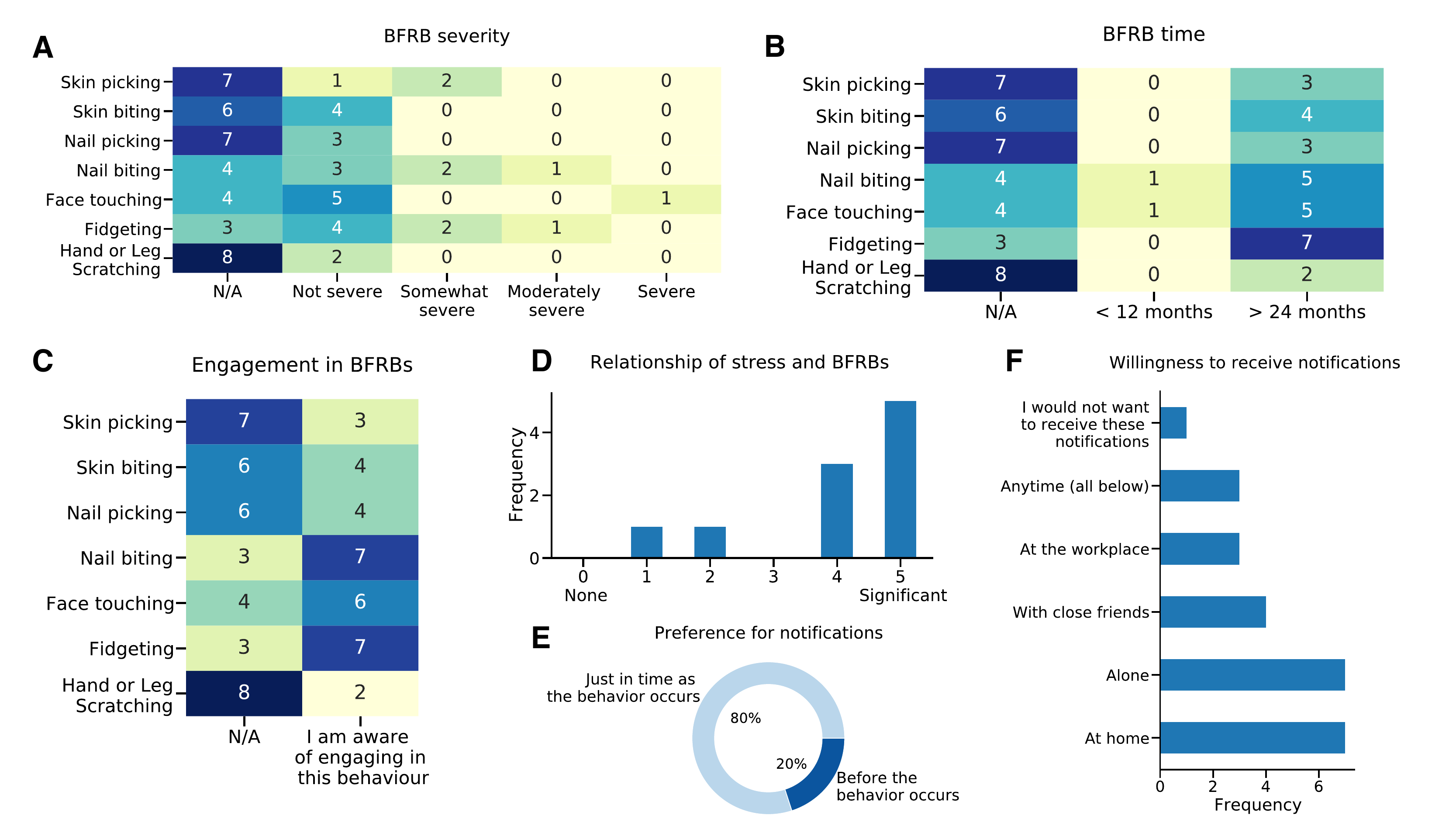}
    \caption{\textbf{Self-reported BFRBs along with perceptions towards stress and potential interventions.} Analysis of a post-experiment follow-up survey in which the participants reported the severity, timeframe and BFRBs' relationship to stress. Attitudes towards the space and time of potential contextual notifications were also recorded. }
    \label{fig:survey}
  \end{figure}


In the quantitative analysis, it was evident that BFRBs prediction can be achieved on both 1-min and 5-min windows; but, the window size prior to the episode matters to the prediction accuracy. To contextualize these findings and understand the role of BFRBs in people's lives, we conducted a follow-up survey with our participants. The survey included two parts. The first part consisted of three questions from the Habit Questionnaire~\cite{teng2002body} that probed BFRBs' presence, severity, and perseverance (Figure~\ref{fig:survey} A--C). The second part consisted of three close-ended questions that probed our participant's stress levels in relation to BFRBs, the willingness to receive notifications via a wearable application, and the preferred timing of receiving such notifications (Figure~\ref{fig:survey} D--F). Additionally, the second part consisted of two open-ended questions: \emph{i) Have any of BFRBs behaviours interfered with your day-to-day activities}, and \emph{ii) Have any of BFRBs behaviours caused injuries or permanent damage?}, which we analyzed through a thematic analysis.




We observed that the severity of these behaviors ranged from `not severe' to `somewhat severe' (Figure~\ref{fig:survey}.A); an expected finding as none of our participants had been diagnosed with these conditions. However, considering that BFRBs are significantly underdiagnosed in the general population \cite{xavier2019improving}, our participants might not be fully aware of the severity of such behaviors, mainly owing to self-reporting bias.  We also observed that almost all behaviors had been prevalent for over 2 years (Figure~\ref{fig:survey}.B), suggesting the prolonged severity of these behaviors.  Interestingly, all but one participant reported willingness to receive notifications about these behaviors (Figure~\ref{fig:survey}.F), suggesting a desire to treat them (or, as a first step, create awareness of these behaviors). This also sheds light onto the scenarios in which participants are willing to receive notifications, with `At home' and `Alone' being the dominant of options. We also found that only 20\% of participants were in favor of the anticipatory notifications, whilst the remaining 80\% would opt for just-in-time results (Figure~\ref{fig:survey}.E). The latter two results offer insights into the desirability of such applications of wearables for consumers, however, it remains unclear whether the results in less-favored scenarios and anticipatory notifications reflect opposition to potentially intrusive notifications, or to social stigmas surrounding BFRBs~\cite{stevenson2018investigation,lochner2013consumer}. Finally, we found that stress was perceived to be a significant cause for BFRB severity and prevalence by our participants (Figure~\ref{fig:survey}.D), with 80\% of results opting for a score of 4 or 5. In conclusion, whilst the severity of behaviors were mostly `not severe', the results reflect a resounding support for personalized applications aiming to reduce the occurrences of these behaviors, even when perceived to be of low severity.

To contextualize our quantitative findings and understand the role of BFRBs in people's lives, we conducted a thematic analysis on the open-ended questions using a combination of open coding and axial coding~\cite{braun2006using}. First, we labeled relevant statements in the two open-ended questions of our survey. Second, axial coding was used to identify relationships between concepts and categories that emerged during open coding. Additional emphasis was given on the role of BFRBs in day-to-day activities, the motivations of doing them, and potential harms/injuries that might have caused. We reviewed themes in a recursive manner than linear, and moved between phases as needed, by repeating and re-evaluating themes and coded text as necessary~\cite{braun2006using}. We found three high-level themes related to: \emph{Context of BFRBs}, \emph{Triggers for BFRBs}, and \emph{Negative Consequences}.\par

\noindent \textbf{Context of BFRBs.} The first most prominent theme concerned the context in which BFRBs occur. Many participants stated workplace-related stressors to be the primary context in which these behaviors are performed. P7 stated that \textit{``During a stressful meeting, I start rubbing/touching face''}, while P3 linked it to a stressful period during exams (\textit{`` I bite my nails in exams and when revising. Face touching is usually done out of boredom but I do rub my forehead in exams.''}. \par

\noindent \textbf{Triggers for BFRBs.} Building upon the theme of context, the second theme concerned triggers that motivate these behaviors. Interestingly, some participants expressed the habitual nature of these behaviors, and not necessarily linking them with stress (or workplace stress). For example, P8 mentioned that \textit{``I touch my face (and beard) on a regular basis and I don’t think it's because of stress or anything. It is the one thing that I constantly do''}. The same participant saw certain behaviors as a way of motivation or distraction from an ongoing task (\textit{``Skin picking and fighting [sic] I don’t do constantly, there are some periods when I do them and some periods where I don’t. It might not be stress-related, but mostly motivation related, as I do them when I feel distracted and (in a sense) don’t want to return back to work. Also I do them while I am more lazy/not-working in other settings (e.g. I am waiting for my turn to get in the shower)''}).\par

\noindent \textbf{Negative consequences.} The third theme concerned potential harms and injuries that these behaviors cause. While most of our participants did not suffer from very severe self-damage or injuries due to these behaviors, some expressed mild negative consequences. P9 stated that \textit{``sometimes he had calluses, but nothing too damaging''}, while P7 had no injuries at all, other than mild \textit{``exacerbating skin issues''}. However, two of our participants stated more severe negative consequences. P2 stated that \textit{``sometimes the based of my fingernails bleeds''}, whereas P4 experienced \textit{``a little inflammation after nail biting''}.\par

\section{Discussion}
\label{sec:discussion}

\subsection{Main results}
While prior research demonstrated the efficacy of detecting compulsive behaviors such as hand-to-mouth and smoking~\cite{azaria2016thumbs,lu2019detection}, there is a dearth of literature in building predictive models to anticipate repetitive behaviors. We conducted a semi-controlled living experiment, and collected a data set comprised of a total of 380 minutes of BFRBs from 10 subjects. By analyzing the collected data, we found that the medical hypotheses (i.e., linked to changes in environment, or stress~\cite{bohne2002}) surrounding these behaviors can be exploited to enhance the accuracy of predictive systems. Higher performances were reported in personalized models, with recalls above 0.8 in certain setups. 1-minute anticipatory window sizes performed best on predicting all-compulsive behaviors, whilst 5-minute models performed consistently throughout each setup when implemented using logistic regression. 

Furthermore, in the 1-minute personalized model, we found that motion sensors perform better when isolated versus heart-rate, while 5-minute models relied more heavily on heart-rate for classification. Similarly, generic models showed strong tendencies to rely on heart-rate, with personalized models gaining performance from motion sensors. Answering our research's overarching goal, we conclude that the prediction of BFRBs can be achieved using limited wearable-sensing data using both 1-minute and 5-minute anticipatory windows. 

In a follow-up survey, our participants reported mostly non-severe BFRBs. However, almost all behaviors had been prevalent, to some extent, for over 2 years, which suggests their prolonged nature. Interestingly, almost all our participants expressed positive attitude towards receiving notifications regarding these behaviors, and some preferred to be notified when they are at home, or alone. This means that a prediction model needs to be ``context-aware'', and offers useful insights into the design of wearable-based interventions (e.g., nudges). However, it remains unclear whether the results in less-favored scenarios and anticipatory notifications reflect opposition to potentially intrusive notifications, or to social stigmas surrounding BFRBs \cite{durna2019public}.\par



\subsection{Implications}
Our work has both theoretical and practical implications. From a theoretical standpoint, our findings concern medical implications. Physiology is a foundational area of medical training and practice, and our work contributes towards the goal of harnessing physiological data to advance clinical machine learning with consumer devices \cite{sarma2020physiology}. For instance, our work can provide insights into diagnosis and disease progression. Whilst reviewing the audiovisual footage, we found there was increased activity (captured from the motion and heart rate sensors) leading up to the compulsive behaviors, confirmed by our results. Based on this, we hypothesize, that compulsive behaviors may differ in intensity and their role in emotion regulation. We foresee that our work would provide an additional dimension to compulsive behavior analysis and diagnosis, by defining the path to clinical diagnosis of BFRBs as progressive in terms of development. 


From a practical perspective, there have been some examples of works dealing with real-time detection of compulsive behaviors relying on motion data and heart-rate data for inference \cite{azaria2016thumbs, lu2019detection}. However, predicting the occurrence of compulsive behaviors ahead of time is a novel area of pervasive health. Our multisensory approach is feasible to be deployed in a real-world application as our models are explainable and lightweight to be ported to wearable devices for continuous monitoring in free-living conditions. The short 1-minute observation windows allow for immediate interventions, while longer windows of 5-minute are more robust in terms of prediction accuracy. Specifically, we found that, heart rate data in conjunction with motion can be used to anticipate compulsive behavior. Motion being a predictive modality was expected due to the nature of behaviors being linked to stressful situations, however, heart rate data shows more promise when leading up to the occurrences of BFRBs. More importantly, the follow-up survey results suggest our participants' willingness to get notified when such behavior(s) occur. Interestingly, the timing of receiving such notifications varied among our participants' responses, suggesting that it plays an important role into how well the notification could be perceived.\par

In general, our BFRBs models could be utilized to support BFRB treatment methods (e.g., CRT~\cite{competingresponse}), create  awareness of such behaviors but, more specifically, enforce them into adopting good personal hygiene practices. Given the recent re-emergence of face-touching as a public health risk for infectious diseases, our face-touching model could be of immediate use. For example, recent preliminary studies found that the use of masks reduced face-touching \cite{shiraly2020face}. In a similar vein, our approach may have a role in COVID-19 transmission slowdown by creating awareness of one's face-touching behavior \cite{perez2020frequency}. As discussed earlier, our models could also complement other modalities such as electrodermal sensors or ultrasound signals emitted by earphones \cite{rojas2021scalable}, towards more comprehensive monitoring of face touching; a prominent use case in light of COVID-19.\par



\subsection{Limitations}
Our work has four main limitations that speak for future work. First, to determine whether our methods could differentiate between the two most prominent behaviors (i.e., skin picking and face touching) an additional experiment was conducted. However, due to the differences in behavior patterns in our data set (Figure \ref{fig:cbperparticipant}), neither cross-validation method provided satisfactory results. Secondly, the missing RMSSD feature from 1-minute windows, due to heart-rate variability measurement traditionally using 5-minute windows \cite{malik1996heart}, also inhibits the analysis when comparing different window sizes. However, more recent research suggests some features may be accurately extracted from smaller windows with promising results \cite{schaaff2013measuring}. We also analyzed different window sizes (2-, 3-, 4-minute), however, the results expectedly decreased as the motion data became noisier whilst still lacking RMSSD data. The third limitation concerns the format of the experiment, resulting in overlapping segments of data from which features were extracted. Finally, the fourth limitation concerns the simulation of real-world use cases. While the user study appropriated real-world scenarios through stress-inducing tasks, our findings should be interpreted under the study's experimental conditions. We acknowledge that presentation and arithmetic tasks might constitute a small fraction of real-world tasks but we think that it fits our set up and study goals. For instance, our study does not account for users on-the-go (e.g., people walking or running), thus is limited to the current experimental setup (i.e., sitting position). Whilst the results indicate our expectations were accurate, future studies would be required to acquire further evidence for the prediction of these compulsive behaviors.\par

To sum up, these limitations primarily arise from the size and labeling of the data set. The length of each behavior is not uniform and is not accounted for in the analysis. The intensity of behaviors is also excluded, but could be used to add additional labels. Further studies would include more participants, potentially even filtering for certain behaviors allowing for more fine-grained tuning. Future work would also extend data gathering to include a free-living study to comprise a larger data set, with greater focus on the potential of personalized models. Scaling the study to support tailored models of participants would be straightforward as calibration only requires a 5-minute baseline reading from the user to begin with. Labeling would be conducted through a combination of both inference and participant feedback to ensure the validity of data gathered from unobserved environments. \par

\section{Conclusion}
\label{sec:conclusion}
Body-focused repetitive behaviors like face-touching or skin-picking are characterized primarily by the use of hands. While there is an abundance of medical evidence on the importance and severity of these behaviors (particularly if not early identified and corrected), there is yet dearth of empirical exploration of these behaviors using consumer-grade wearables. We conducted a feasibility study in which participants were exposed to a series of tasks that induced BFRBs. By analyzing a total of 380 minutes of raw signals under an extensive evaluation of sensor modalities, cross-validation methods, observation windows, and machine learning classifiers, we found that BFRBs can be predicted with sufficient accuracy. In particular, generic compulsive behavior (vs. normal) achieves 89.2 AUC, face touching 94.4 AUC, and skin picking 94 AUC. Across most models, using an observation window of 5 minutes prior to the episode outperformed the 1 minute one and, notably, the heart sensors were stronger predictors in the former while the motion sensors dominated in the latter. These findings are a fundamental step towards creating awareness of one's BFRBs (given the recent re-emergence of face touching as a public health risk for infectious diseases), and designing just-in-time interventions to prevent them.

\begin{acks}
This work is partially supported by Nokia Bell Labs through their donation to the Centre of Mobile, Wearable Systems and Augmented Intelligence at the University of Cambridge. D.S is additionally supported by the Embiricos Trust Scholarship of Jesus College Cambridge, and the EPSRC through Grant DTP (EP/N509620/1). The authors declare that there is no conflict of interest regarding the publication of this work.
\end{acks} 

\bibliographystyle{ACM-Reference-Format}
\bibliography{references}


\end{document}